\documentclass[preprint,12pt]{elsarticle}

\usepackage{amssymb}
\usepackage{color}
\usepackage{amsbsy}
\usepackage{amsmath}
 \newcommand{\comments}[1]{}   %%%%%%%%%%%%%%% comments

\journal{Chaos, Solitons \& Fractals}

\begin{document}

\begin{frontmatter}



\title{Fluctuations of rotational and translational degrees of freedom in an interacting active dumbbell system}


\author[label1]{Leticia F. Cugliandolo}
\ead{leticia@lpthe.jussieu.fr}

\author[label2]{Giuseppe Gonnella}
\ead{gonnella@ba.infn.it}

\author[label3]{Antonio Suma}
\ead{antonio.suma@gmail.com}

\address[label1]{Sorbonne Universit\'es, Universit\'e Pierre et Marie Curie - Paris VI, \\
Laboratoire de
Physique Th\'eorique et Hautes \'Energies, \\
4 Place Jussieu, 75252 Paris Cedex 05,
France}

\address[label2]{Dipartimento di Fisica, Universit\`a di Bari {\rm and}  \\
INFN, Sezione di Bari, via Amendola 173, Bari, I-70126,
Italy}

\address[label3]{SISSA - Scuola Internazionale Superiore di Studi Avanzati,\\
Via Bonomea 265, 34136 Trieste 
Italy}

\begin{abstract}

%\textcolor{red}{To be shorten}
We study the dynamical properties of a two-dimensional ensemble 
of self--propelled dumbbells with only repulsive interactions. 
After summarizing the behavior of the translational and rotational mean-square displacements  in the homogeneous phase
that we established in a previous study, we analyze their fluctuations. We study the dependence
of the probability distribution functions  in terms of the 
P\'eclet number, describing the relative role of active forces and thermal fluctuations,  
and of particle density. 
%\\
%We find that the four distinct regimes of the translational mean square displacement of the 
%single active dumbbell survive at finite density for parameters that lead to a separation of time-scales.
%We establish the Pe and density dependence of the diffusion constant in the 
%last diffusive regime. We prove that the ratio between the  diffusion constant and its value for the single dumbbell
%depends on temperature and active force only 
%through the P\'eclet number at all densities explored.
%We also study the  rotational mean square displacement  proving the existence of a rich 
%behavior with intermediate  regimes only appearing at finite density. 
%The ratio of the rotational late-time diffusion constant and  its vanishing density limit
%depends on Pe and density only. At low P\'eclet number it is a monotonically decreasing function of density. At
%high P\'eclet number it first increases to reach a maximum and next decreases as a function of density.

\end{abstract}


\end{frontmatter}


\section{Introduction}
\label{sec:introduction}

Active matter is characterised by the continuous partial conversion of  internal energy into work. 
Some examples, at different scales, are the cytoskeleton, 
bacterial colonies, algae suspensions,  bird flocks and schools of
fish. Self-propelled units  can also be artificially realized in the laboratory in 
different ways, for example, by surface treatment of colloidal particles~\cite{Golestanian05,Howse07}. All these systems live, or function,
 in conditions far from thermodynamic equilibrium  and pose challenging questions to non-equilibrium statistical mechanics.
%The diffusion properties of a suspension are significantly affected by self-propulsion. 
Active matter exhibits non-trivial collective properties
that have no analogue in passive  materials such as large scale coherent motion  
in the absence of any attractive 
interaction
%~\cite{Mendelson99,wu,Dombrowski04,Hernandez05,Riedel05,Peruani06,Sokolov07,Zhang09},
and a phase separation into an aggregate and a gas-like phase.
%~\cite{Tailleur08,Fily12,Fily14,Redner13,Stenhammer13,Suma13,Suma14,Levis-Berthier,Stenhammer14,Buttinoni13b}.  
Several review articles are devoted to this rapidly developing field of research~\cite{Toner05,Fletcher09,menon,Ramaswamy10,Cates12,Romanczuk12,Vicsek12,Marchetti13,Marenduzzo14,Elgeti15,gonn15}. 

The diffusive properties in these systems are of particular interest.
 A number of experimental and numerical studies addressed how the diffusive properties are affected by self-propulsion and the 
density of the suspension; some  focused on the dynamics of {\it passive tracers} immersed in the active bath~\cite{wu}, others
focused instead on the mean-square displacement of the {\it active particles} 
themselves~\cite{Hernandez05,Mino11,Gregoire01,Llopis06}.

An interesting model of active matter is one in which the active components have the
elongated shape of many natural swimmers.
A first study of the phase diagram of such a system with active  dumbbells molecules~\cite{valeriani2011colloids} 
appeared  in~\cite{Suma13,Suma14}. The analysis of 
effective temperature ideas~\cite{cugl:review}, and the averaged rotational and translational mean-square 
displacements were presented in~\cite{Suma14b} and \cite{Suma14c}, respectively, for a two-dimensional system.

In this paper we recall some of the results in these publications and we extend the analysis to the 
fluctuations of translational and rotational degrees of freedom. 
%We then extend the 
%analysis to the distribution functions of translational and rotational displacements over fixed 
%time-delays falling in different time-regimes already identified in the study of the translational and 
%angular mean-square displacements. 
%The paper is organised as follows. 
In Section 2 the dumbbell model is very briefly explained. 
In Section 3 the numerical results for the translational and rotational fluctuations in the 
interacting active system are presented. A discussion will complete the paper in Section 4.

\section{The model}
\label{sec:model-dumbbells}

We briefly present the model and the parameters used in the simulations. 
More details can be found in~\cite{Suma14b,Suma14c}. 
The dumbbells are diatomic molecules formed by two spherical colloids, 
elastically linked together {\it via} the finite extensible non-linear elastic force
\begin{equation}
{\mathbf F}_{\rm fene} = - \frac{k {\mathbf r}}{1- (r^2/r_0^2)} 
\; , 
\end{equation}
with $k>0$ and ${\mathbf r} = {\mathbf r}_1 - {\mathbf r}_2$
the vector linking the centres of the spherical colloids, with diameter $\sigma_{\rm d}$ 
and mass $m_{\rm d}$.  An additional Weeks-Chandler-Anderson potential,
\begin{eqnarray}
\label{eq:WCA-potential}
V_{\rm wca}( r ) 
&=&
\left\{
\begin{array}{ll}
V_{\rm LJ}( r ) - V_{\rm LJ}(r_c) & \qquad r<r_c
\nonumber\\
0 & \qquad r > r_c
\end{array}
\right.
\end{eqnarray}
with 
\begin{equation}
V_{\rm LJ}(  r  ) = 4\epsilon \left[ \left( \frac{\sigma_{\rm d}}{r} \right)^{12} - \left( \frac{\sigma_{\rm d}}{r} \right)^{6}\right]
\; ,
\end{equation}
where $\epsilon$ is an energy scale and $r_c$ is the minimum of the Lennard-Jones potential, $r_c=2^{1/6} \sigma_{\rm d}$, 
is added to ensure that the colloids in the same molecule do not overlap.
The active forces  are polar and act along the main molecular axis  $\hat {\mathbf n}$, are constant in modulus but follow the 
molecules' rotations, and are the same for the two spheres belonging to the same molecule, 
\begin{equation}
{\mathbf F}_{\rm act} = F_{\rm act} \ \hat {\mathbf n}
\; . 
\end{equation}
${\mathbf F}_{\rm act}$ is  directed 
from the $i$th colloid (tail) to the $i+1$th colloid (head). The active forces are  applied to all molecules
in the sample during all their dynamic evolution. 
We take the interaction between the spheres in different dumbbells to be purely repulsive.
 
Putting these ingredients together, the dynamic equations are
\begin{eqnarray}
&& m_{\rm d}\ddot{{\mathbf r}}_{i}(t) \! = \! -\gamma \dot{{\mathbf r}}_{i}(t)+
 {\mathbf F}_{\rm fene}({\mathbf r}_{i, i+1})+
 {\boldsymbol \eta}_{i} 
% &&
 - \!
 \sum_{\substack{
            j=0\\
            j \neq i}}^{2N} 
 \frac{\partial V_{\rm wca}^{ij}}{\partial  r_{ij}}
 \frac{{\mathbf r}_{ij}}{ r_{ij}}+ {{\mathbf F}_{\rm act}}_i
\; , 
\label{eqdumbattcoll}
\\
&& m_{\rm d}\ddot{{\mathbf r}}_{i+1}(t) \! = \!
 -\gamma \dot{{\mathbf r}}_{i+1}(t)-
 {\mathbf F}_{\rm fene}({\mathbf r}_{i,i+1})+
 {\boldsymbol \eta}_{i+1}
% \nonumber\\
% &&
 - \!\!\! \sum_{\substack{
            j=0\\
            j \neq i+1}}^{2N} 
\!\!  \frac{\partial V_{\rm wca}^{i+1,j}}{\partial r_{i+1,j}}
 \frac{{\mathbf r}_{i+1,j}}{r_{i+1,j}}+{{\mathbf F}_{\rm act}}_i  
 \;\;\;\;
\nonumber
\end{eqnarray}
with $i=1,3,...2N-1$,  ${\mathbf r}_{ij} = {\mathbf r}_i - {\mathbf r}_j$, $r_{ij} = |{\mathbf r}_{ij}|$
 and $V_{\rm wca}^{ij} \equiv
V_{\rm wca}(r_{ij})$ with $V_{\rm wca}$ defined in Eq.~(\ref{eq:WCA-potential}).

The coupling to the thermal bath at temperature $T$ is modeled in the manner of Langevin, with 
$\gamma$ the friction coefficient and ${\boldsymbol \eta}$ a Gaussian random noise  with zero 
mean, $\langle \eta_{ia}(t) \rangle = 0$, and
\begin{eqnarray} 
\langle \eta_{ia}(t) \eta_{jb}(t') \rangle &=& 2 \gamma k_BT \delta_{ij} \delta_{ab} \delta(t-t')
\; ,
\label{eq:noise-corr}
\end{eqnarray}
with $k_B$ the Boltzmann constant. $a$ and $b$ label the coordinates in $d$ dimensional space. An effective 
rotational motion is generated by the random torque due to the white noise acting independently on the two beads.

The surface fraction is 
\begin{equation}
\phi=N \ \frac{S_{\rm d}}{S} 
\;  
\end{equation}
with $S_{\rm d}=\pi \sigma_{\rm d}^2/2$ the area occupied by an individual dumbbell in $d=2$, $S$ the total area of the box  and $N$ their total number. 
The spring is supposed to be massless and void of surface.  We impose periodic boundary conditions on the two directions. 

%\begin{figure}[h]
%\includegraphics[scale=0.3]{fig1}
%\cdaption{(Color online.) A sketch of an active dumbbell molecule.}
%\label{fig:dumbbell-act}
%\end{figure}

The P\'eclet number, ${\rm Pe}$, is a dimensionless ratio between the  
advective transport rate and the diffusive transport rate.
For particle flow one defines it as 
%\begin{equation}
$
%{\rm Pe} = \frac{Lv}{D} 
{\rm Pe} = Lv/{D} 
$,
%\;
%\end{equation}
with $L$ a typical length, $v$ a typical velocity, and $D$ a typical diffusion constant. We choose $L \to \sigma_{\rm d}$, $v\to F_{\rm act}/\gamma$ and 
$D\to D^{\rm pd}_{\rm cm}= k_BT/(2\gamma)$ of the passive dumbbell to be derived below; then,
\begin{equation}
{\rm Pe} = \frac{2\sigma_{\rm d} F_{\rm act}}{k_BT}
\; . 
\label{eq:Peclet}
\end{equation}
The active Reynolds number 
$
{\rm Re}_{\rm act} 
= m_{\rm d} F_{\rm act}/ (\sigma_{\rm d} \gamma^2) 
$
is defined in analogy with the hydrodynamic Reynolds number.
% ${\rm Re}=  L v /\nu$, where $\nu $ is the kinematic viscosity of a given fluid, representing the ratio between inertial
% and viscous forces. Here we set $L \to \sigma_{\rm d}$, $v\to F_{\rm act}/\gamma$ and 
% $\nu \to \gamma \sigma_{\rm d}^2 /m_{\rm d}$.

\section{Single dumbbell dynamics}
\label{sec:single-dumbbell-dynamics}
%\subsection{A single dumbbell}

The averaged single dumbbell motion can be derived analytically under the hypothesis that $r\approx \sigma_{\rm d}$. 
Details on the
calculations can be found in~\cite{Suma14b,Suma14c}. Here, we simply summarise the main results. Within this approximation, 
at absolute times and time-differences
that are longer than the inertial time-scale $t_I = m_{\rm d}/{\gamma}$, not taking into account its periodic character, 
the angle $\theta$ between the dumbbell's main molecular axis and an axis fixed to the laboratory is a 
Gaussian random variable with mean $\langle \theta \rangle = \theta_0$ that 
diffuses according to 
\begin{equation}
\langle \theta^2\rangle = \theta_0^2 + 2 D_R t 
\end{equation}
with $\theta_0$ the initial angle, $t$ the time-delay, and the angular diffusion constant
\begin{equation}
\qquad D_R =  \frac{2 k_BT}{\gamma \sigma_{\rm d}^2}
\label{eq:DR}
\; . 
\end{equation}
Averaging over the initial angles, taken from a flat distribution around $\theta_0=0$, yields 
$\langle \theta \rangle =0$ and, in the long times limit, $\langle \theta^2\rangle \to 2D_R t$. 
In the absence of interactions, the angular displacements, $\Delta \theta$ between two 
times that are longer than $t_I$ is also Gaussian distributed.
If one imposes the
periodicity of the angles in the interval $[-\pi, \pi]$ the angular distribution becomes flat.

The translational mean-square displacement is ballistic in the limit $t \ll t_I$, and crosses over to a very rich behaviour 
beyond this time-scale,
\begin{equation}
 \langle\triangle{\mathbf r}_{\rm cm}^2\rangle (t)=4D_{\rm cm}^{\rm pd} \ t+
 \bigg(\frac{F_{\rm act}}{\gamma}\bigg)^2\frac{2}{D_R}\left(t-\frac{1-e^{-D_Rt}}{D_R}\right)
 \; ,
 \label{msddumbattivasingola}
\end{equation}
where 
\begin{equation}
D_{\rm cm}^{\rm pd}=\frac{k_BT}{2\gamma}
\end{equation} 
is the diffusion constant  in the passive
limit, ${\mathbf F}_{\rm act}=0$, see \cite{Suma14b}.
This equation presents several time scales and limits. For $t_I \ll t \ll t_a=D_R^{-1}$ one finds
\begin{equation}
 \langle\triangle{\mathbf r}_{\rm cm}^2\rangle=4D_{\rm cm}^{\rm pd} \ t+
 \bigg(\frac{F_{\rm act}}{\gamma}\bigg)^2t^2 \; ,
 \label{msddumbattivasingola_1limite}
\end{equation}
that can still be split into the passive
diffusive limit $\langle\triangle{\mathbf r}_{\rm cm}^2\rangle=4D_{\rm cm}^{\rm pd} \ t$ for $t_I\ll t<t^* \ll t_a$, and a ballistic
regime  $\langle\triangle{\mathbf r}_{\rm cm}^2\rangle=(F_{\rm act}/\gamma )^2 \ t^2$
for $t^*<t \ll t_a$, where the time scale $t^*$ is given by 
\begin{equation}
t^*=\frac{4D_{\rm cm}^{\rm pd}\gamma^2}{F_{\rm act}^2} = \frac{2k_BT \gamma}{F_{\rm act}^2} 
= \left(\frac{4}{\mbox{Pe}}\right)^2 \ \frac{\sigma_{\rm d}^2}{4D_{\rm cm}^{\rm pd}} = \left(\frac{4}{\mbox{Pe}}\right)^2 \ t_a
\; . 
\end{equation}
Note that these two intermediate 
regimes might be hidden if the system parameters are such that
$t^*<t_I$ or $t^*>t_a$. They can also be easily confused with a super-diffusion behavior
 $ \langle\triangle{\mathbf r}_{\rm cm}^2\rangle \sim   t^\alpha$ 
with $1 < \alpha < 2$ if they are not well separated ($t_I \simeq t^* \simeq t_a$). 
In the large Pe limit one has $t^* \ll t_a$.
 In the last time-lag regime $t\gg t_a$, we recover normal diffusion,
 \begin{equation}
 \langle\triangle{\mathbf r}_{\rm cm}^2\rangle=4D_A \ t \; ,
 \label{msddumbattivasingola2_limite}
\end{equation} 
with the diffusion coefficient
\begin{equation}
D_A(F_{\rm act}, T, \phi=0) = \frac{k_BT}{2\gamma} \left[ 1+ \frac{1}{2} \left(\frac{F_{\rm act} \sigma_{\rm d}}{k_BT}\right)^2 \right]
= D_{\rm cm}^{\rm pd} \ \left(1+ \frac{\mbox{Pe}^2}{8}\right)
\; . 
\label{eq:DA}
\end{equation}

\subsection{The distributions for the single dumbbell}
\label{subsec:distribution-dumbbell}

%\textcolor{blue}{ho modificato gli intervalli per ricavare $p_n_x$. Per ogni soluzione tra $-1<n_x<1$ ce ne sono due, una tra 0,pi e l'altra tra -pi,0, quindi l'intervallo da fissare e' per  theta non per nx}
The infinitesimal increment of the centre of mass position, ${\rm d}x_{\rm cm}$, is a random variable and its distribution 
is due to the Gaussian random character of the noise
$\xi_x$ and the, for the moment unknown, random character of $n_x \equiv \cos\theta$.
In order to establish the pdf of the latter, we assume that $\theta$ is uniformly 
distributed in the interval $[-\pi,\pi]$, $p_\theta(\theta)=(2\pi)^{-1}$. Therefore, for $0 \leq \theta \leq \pi$
\begin{equation}
p_{n_x}(n_x) =  \int_{0}^{\pi} \frac{{\rm d}\theta}{2\pi} \ \delta(n_x-\cos\theta) 
%=  \int_{n_x-1}^{n_x} \!\! \frac{1}{2\pi} \frac{{\rm d} z \ \delta(z)}{\sqrt{1-(n_x-z)^2} }
= \frac{1}{2\pi\sqrt{1-n_x^2} } 
\end{equation}
This calculation can be repeated for $-\pi \leq \theta <  0$ with the same result and the overall 
numerical pre-factor is fixed by normalisation:
\begin{equation}
p_{n_x}(n_x) = \frac{1}{\pi \sqrt{1-n_x^2} } 
\qquad\quad
\mbox{for} 
\qquad n_x \in [-1,1] \; . 
\end{equation}

In order to estimate the pdf of the translational displacement of the centre of mass position, 
we now use the over-damped equation for a small increment 
${\rm \Delta}x_{\rm cm}$, with the notation used in~\cite{Suma14c}
\begin{equation}
{\rm \Delta} x_{\rm cm}=\frac{\sqrt{4k_BT\gamma}}{2\gamma}\ {\rm \Delta}W +v_{\rm act} {\rm \Delta} t \ \cos\theta
\end{equation}
%with $\xi_x$ a Gaussian random variable with $\langle \xi_x\rangle=0$ and 
%$\langle \xi^2_x\rangle = 4\gamma k_BT \ ({\rm d} t)^{-1}$ and $v_{\rm act}=F_{\rm act}/\gamma$.
%{\rm d} x_{\rm cm}=\frac{\sqrt{4k_BT\gamma}}{(2\gamma)}\ dW_{dt}+v_{\rm act} {\rm d} t \ \cos\theta
%\end{equation}
with $v_{\rm act}=F_{\rm act}/\gamma$ and ${\rm \Delta}W$  a gaussian noise, distributed as
\begin{equation}
p_{{\rm \Delta}W}(\zeta)=\frac{1}{\sqrt{2\pi {\rm \Delta}t}} \ e^{-\frac{\zeta^2}{2{\rm \Delta}t}} \; ,  
\end{equation}
with $\zeta\in [-\infty,\infty]$. 
\comments{
\textcolor{green}{Cosi come scritta la $\zeta$ ha le stesse proprieta' della $\eta$ della sezine precedente 
ed il suo quadrato e' infinito. Riporto qualche osservazione dal libro di Zinn-Justin. E' scritto che il rumore
gaussiano $\nu$ delta correlato e' distribuito con $\exp {\frac{-1}{2 \Omega}\int{\nu^2}dt }$. Poi ZJ nel capitolo sulle eq stocastiche
fa i conti sempre con la distribuzione che ho appnea scritto lasciando l'integrale. C'e' pero' un appendice al capitolo su discretized Langevin equation. Qui c'e' scritto che nel limite di small time step vale la distribuzione gaussiana che avete scritto voi.  ma credo basti solo cambiare qualche parola in quello che abbiamo scritto. Possiamo dire invece di infinitesimo, small increment o qualcosa del genere? Poi direi  che dw
e' distribuito con una gaussiana  con varianza =... Infine, citare Ito potrebbe confondere perche' nell'altro lavoro 
utiizziamo Stratonovich. Che dite? Per il resto toglierei tutti i colori e Antonio, hai controllato i fattori della 22?}
\textcolor{blue}{ho sostituito gli infinitesimi con intervalli finiti, perché in effetti la formula 19 e' valida solo se integri in un intervallo infinitesimo l'equazione del moto e fai una approssimazione di eulero (gardiner 10.1.1)}
}

 The pdf of ${\rm \Delta}x_{\rm cm}$ is then
 \begin{eqnarray*}
 \;\;\;\; p({\rm \Delta}x_{\rm cm})  \! = \! 
 \int {\rm d}\zeta {\rm d} n_x \ 
 \delta\!\left({\rm \Delta} x_{\rm cm} - \frac{\sqrt{4k_BT\gamma}}{2\gamma} \zeta - v_{\rm act} {\rm \Delta} t \ n_x \! \right) 
 p_{n_x}(n_x) p_{\Delta W}(\zeta)
 \end{eqnarray*}
 The naive comparison of the order of magnitude of the last two 
 terms inside the delta-function using $\zeta \sim  ({\rm \Delta}t)^{1/2}$
 and $n_x\simeq 1$ yields, after multiplying by $\sigma_d\gamma/(k_BT\Delta t)$, 
 \begin{equation}
 \left(\frac{\sigma^2_{\rm d} \gamma}{k_BT{\rm \Delta}t}\right)^{1/2} \approx \frac{\sigma_{\rm d} F_{\rm act} }{k_BT}
 \; . 
 \label{eq:scale}
 \end{equation}
In this way the second member corresponds to Pe.
 Two limits are clear: \\
 -- For Pe $\ll $ left-hand-side in Eq.~(\ref{eq:scale}), so that the active-force induced last term in the argument of the Dirac delta can be neglected, the two integrals decouple and 
 ${\rm \Delta}x_{\rm cm}$ is naturally Gaussian distributed. 
 \\
-- For Pe $\gg $ left-hand-side in Eq.~(\ref{eq:scale}), so that the noise term in the argument of the Dirac delta can be neglected, the two integrals decouple again and 
\begin{equation}
 p({\rm \Delta}x_{\rm cm}) \propto 
 \frac{1}{\sqrt{1- \left( \frac{{\rm \Delta}x_{\rm cm}}{v_{\rm act} {\rm \Delta} t} \right)^2 }}
\end{equation}
with two peaks at ${\rm \Delta}x_{\rm cm} = v_{\rm act} {\rm \Delta} t$.
\\
Otherwise, the double integral yields a complex result:
\begin{eqnarray}
p({\rm \Delta} x_{\rm cm}) \propto 
\int_{-1}^1 \!\! {\rm d} n_x  \exp\left[ - \frac{(2\gamma)^2}{2{\rm \Delta}t (4k_BT\gamma)} \  ({\rm \Delta} x_{\rm cm} + v_{\rm act} {\rm \Delta}t \ n_x)^2 \right] \frac{1}{\sqrt{1-n_x^2}}
\label{eq:pdf-dxcm}
\end{eqnarray}

The displacement between two times, $\Delta x = x_{\rm cm}(t+t_0) - x_{\rm cm}(t_0)$,
is the sum of the small increments with each of these 
independently distributed according to Eq.~(\ref{eq:pdf-dxcm}).
%,  $P(\Delta x) = \int \prod_{i=1}^N {\rm d}x_{\rm cm}^i \
%p({\rm d}x_{\rm cm}^i) \ \delta(\Delta x - \sum_{i=1}^N {\rm d} x^i_{\rm cm}) $.
%\textcolor{green}{quest'ultima formula la toglierei, dicendo a parole, poi vediamo} \textcolor{red}{la lascerei, perche 
%serve a definire $\Delta x$, senza el $_{\rm cm}$, come lo usiamo nelle figure})

\section{Translational motion at finite density}

\begin{figure}[!h]
\begin{center}
  \begin{tabular}{ccc}
  \hspace{-2cm}
      %\includegraphics[scale=1]{Fa0_1T0_005fi0_5.eps}
      %\hspace{-3.2cm}
      %\includegraphics[scale=1]{Fa0_1T0_005fi0_7.eps}
      %\hspace{0.5cm}
      %\\
      %\hspace{-2cm}
      \includegraphics[scale=1]{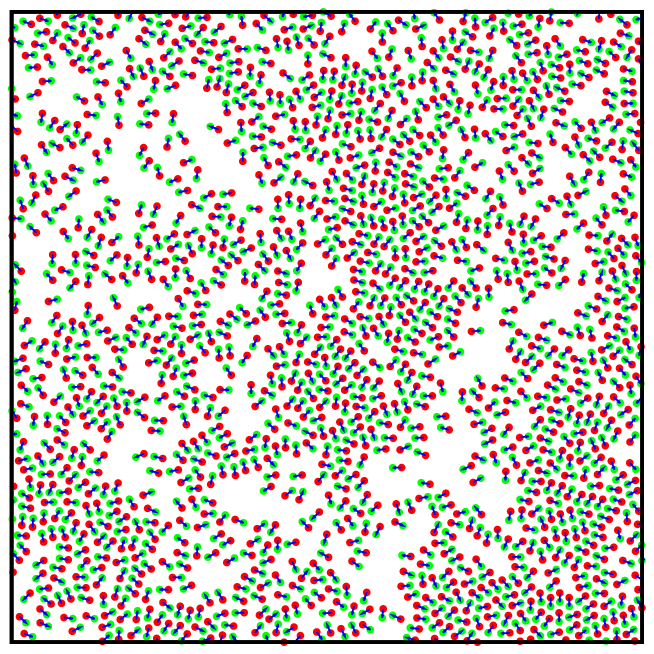}
      \hspace{-3.2cm}
      \includegraphics[scale=1]{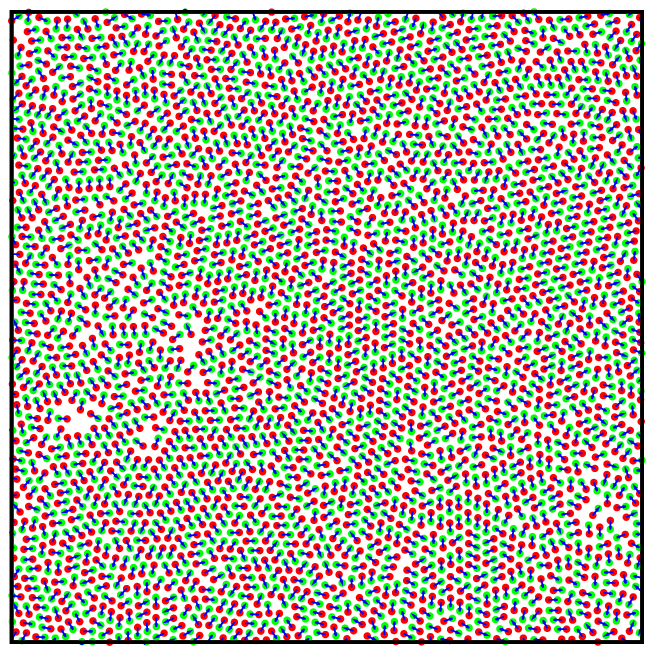}
      \hspace{0.5cm}
    \end{tabular}
\caption{Typical snapshots of a system with $\phi= 0.5, \ 0.7$, linear size $70 \ \sigma_{\rm d}$ and Pe = $40$. In each dumbbell the 
green colloid is the head and the red one is the tail, and  they are joined by a line. }
\label{fig:snapshots} 
\end{center}
\end{figure}

The interacting case cannot be solved analytically. We focus here on the numerical determination  of
the centre of mass and angular displacement statistics.
Details on the numerical method used for solving the dynamical equations (\ref{eqdumbattcoll})
are given in \cite{Suma14b}. We set $m_{\rm d} = \sigma_{\rm d} = k_B = \epsilon = 1$,  
$r_0 = 1.5$, $k =  30$ and $\gamma=10$ in proper physical units. These choices   assure over-damped motion and 
negligible dumbbell vibrations. 
We used between 15000 and 20000 dumbbells in each simulation. We fix the strength of the 
active force to be $F_{\rm act}=0.1$ and we vary the temperature in order to access different P\'eclet numbers
that we choose to be Pe = 40, 20, 4, 2. The remaining parameter is density and we typically use $\phi=0.01, \ 
0.1, \ 0.3, \ 0.5,  \ 0.7$, see Fig.~\ref{fig:snapshots} that shows two instantaneous snapshots of the dumbbell configurations
of systems with $\phi=0.5$ (left) and $\phi=0.7$ (right). The single-dumbbell characteristic time-scales for these parameters are summarised in 
Table~\ref{table-uno}.

\vspace{0.75cm}

\begin{table}[h]
\centerline{
\begin{tabular}{| l | r | r | r  | c | }
\hline
$T$ & Pe & $t_a$ & $t^*$ 
\\
\hline
\hline
0.005 & 40 & 1000 & 10
\\
0.01 & 20 & 500 & 40
\\  
0.05 & 4  & 100 & 1000
\\ 
0.1 & 2 & 50 & 4000
\\
\hline
\end{tabular}
}
\caption{Parameters and characteristic times as defined in Sect.~\ref{sec:single-dumbbell-dynamics}.}
\label{table-uno}
\end{table}

Aspects of the phase diagram and the dynamics of this system were already established in~\cite{Suma13,Suma14,Suma14b}. 
At sufficiently low temperature and large active force the system phase separates into gas-like spatial regions and clusters of 
agglomerated dumbbells. The dynamic phase transition between homogeneous and aggregated phases was determined 
by the change in behaviour of the probability distribution function, $\rho$, 
of the local density, $\phi_x$~\cite{Suma13,Suma14,Suma14b}.
At the critical Pe at which the system starts aggregating 
the density distribution $\rho$  not only becomes asymmetric but starts
developing a second peak at the density of the clusters. Snapshots of typical  configurations 
and their analysis along these lines can be found in~\cite{Suma14c}. 
In the rest of the paper we use sufficiently low P\'eclet numbers  so  that the system 
is in the homogenous phase though with important fluctuations, 
as we will see.

\subsection{Center of mass translational mean-square displacement} 

\begin{figure}[!h]
\begin{center}
  \begin{tabular}{cc}
      \includegraphics[scale=0.67]{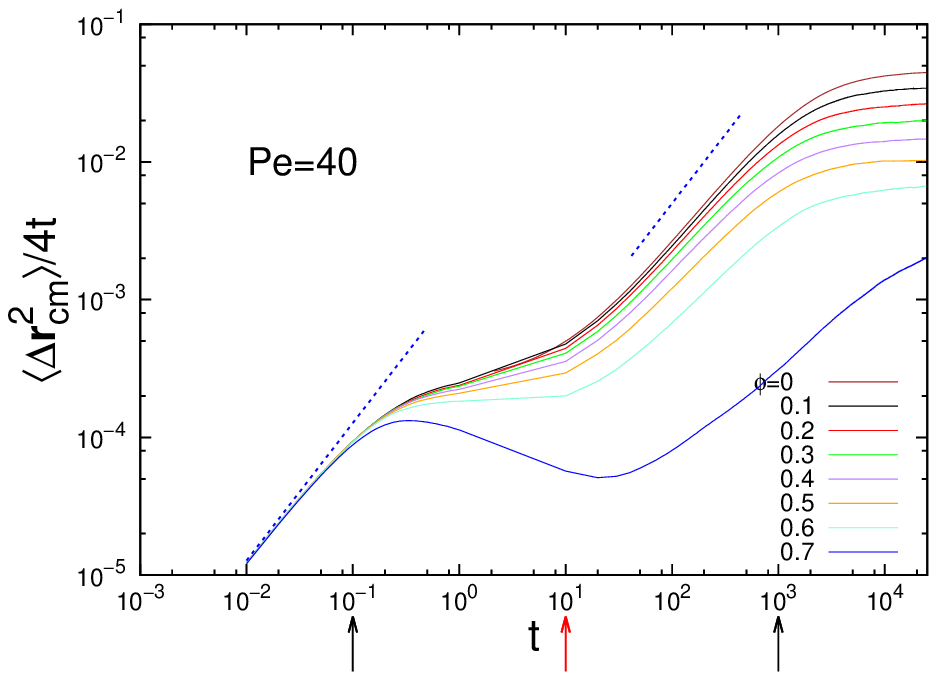}
          \includegraphics[scale=0.67]{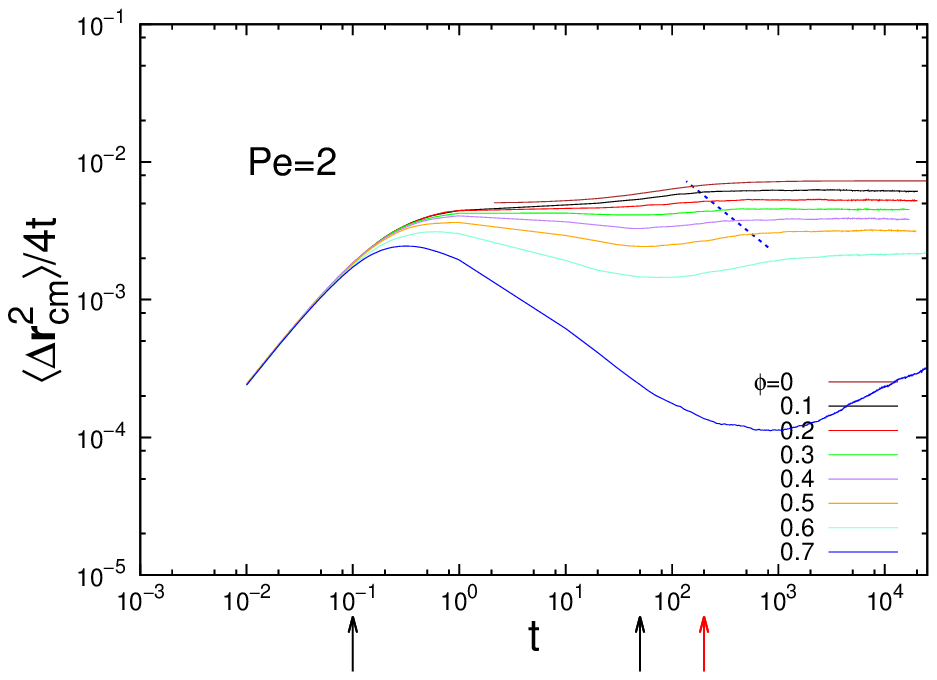}    
    \end{tabular}
\caption{The center of mass MSD normalised by time-delay, for an active system 
under $F_{\rm act}=0.1$ at $T=0.005, \ 0.1$ (Pe = $40, \ 2$), 
with different densities given in the key.
The P\'eclet number induces a strong qualitative change in $\langle \Delta {\mathbf r}_{\rm cm}^2 \rangle$, 
see the text for a detailed discussion. The two dashes in the first panel represent the ballistic dependence $\simeq t^2$. 
The dashed segment in the last panel is a guide-to-the-eye for the density dependence of the last crossover time-delay
that increases weakly with $\phi$. 
The vertical black arrows indicate the single dumbbell time-scales $t_I$ and $t_a$, while the  red arrows
indicate the single dumbbell characteristic time $t^*$, for each case. 
}
\label{fig:mean_square_displacement_cm} 
\end{center}
\end{figure}

In Fig.~\ref{fig:mean_square_displacement_cm} we show  the center of mass Mean Square Displacement (MSD) 
normalised by time-delay in such a way that a plateau signals normal diffusion. The two panels 
display data at $T=0.005$ and $T=0.1$, under the same active force $F_{\rm act} =0.1$. Each panel shows data for eight
densities given in the key and the case $\phi=0$ corresponds to the single dumbbell problem.  
The characteristic times $t_I, \ t^*, \ t_a$ (see Table~\ref{table-uno}) are shown with small vertical arrows. 
These plots show several interesting features that reproduce, to a certain extent, the single particle motion summarised above:\\
-- In all cases there is a first ballistic regime (the dashed segment close to the data is a guide-to-the-eye) 
with a pre-factor that is independent of $\phi$ and increases with temperature. 
(The case $t\ll t_I$ of the single  dumbbell.)
\\
-- The dynamics slow down next and, depending on Pe and $\phi$, the normalised MSD attains a 
plateau associated to normal diffusion or decreases, suggesting sub-diffusion.
(The case $t_I \ll t \ll t^* \ll t_a$ of the single dumbbell.) For instance, there is sub diffusion for Pe = 2 and 
$\phi \geq 0.4$ and Pe = 40 and $\phi \geq 0.7$. 
\\
-- Subsequently, the dynamics accelerate with a second super-diffusive regime in which the curves 
for all $\phi$ look approximately parallel and very close to ballistic at sufficient high Pe.
%$T=0.005, \ 0.01, \ 0.05$. 
(The case $t_I \ll t^* \ll t \ll t_a$ of the single dumbbell.) 
\\
-- Finally, the late normal diffusive regime is reached with all curves saturating at $D_A$.
(The case $t_I \ll t^*, \ t_a \ll t$  of the single dumbbell.) This regime is beyond the time-window 
explored for $\phi = 0.7$.

It is hard to ensure whether  the intermediate regime is super-diffusive or simply ballistic as the time-scales 
$t^*$ and $t_a$ are not sufficiently well separated at high Pe and not even ordered as $t^*< t_a$ 
at low Pe, leading to the mixture of the diffusion-ballistic-diffusion regimes.
%are mixed, due to the fact that the condition $t^* \ll t_a$ is no longer satisfied.  
%The  effective slope in the intermediate  super-diffusive regime
%decreases when the density increases, \textcolor{green}{as it is better seen at high Pe}. \textcolor{red}{not very sure 
%about this claim, look at the Pe = 2 plot, the increases seems more abrupt. Numerical error?}
 
 A rather good fit, not shown here, 
of the finite density data in the limit Pe $\gg 1$ and for time-delays such that $t \ge t^{*}$
 is achieved by using the single dumbbell  expression in Eq.~(\ref{msddumbattivasingola}) 
  \begin{equation}
 \langle\triangle{\mathbf r}_{\rm cm}^2\rangle (t)=
 4D_A^{\phi}\bigg(t-\frac{1-e^{-D_R^{\phi}t}}{D_R^{\phi}}\bigg)
 \; ,
 \qquad\qquad\quad \mbox{Pe} \gg 1
 \; ,
 \label{fitmsddumbattivasingola}
\end{equation}
without the first term (negligible if Pe $\gg 1$) and upgrading the remaining parameters, $D_A^{\phi}$ and $D_R^{\phi}$, 
to be density-dependent fitting parameters, as  done in~\cite{wu,Fily12}. The quality of this fit was discussed in~\cite{Suma14c}.
\comments{ %%%%%%%%%%%%% description of fit 
is shown in
Fig.~\ref{fig:mean_square_displacement_cm-fit} (left panel). For not that large values of Pe
one could recover the remaining parameter and use instead
$\langle\triangle{\mathbf r}_{\rm cm}^2\rangle (t)=
 4D^{{\rm pd}, \phi}_{\rm cm} t +4D_A^{\phi} (t-\frac{1-e^{-D_R^{\phi}t}}{D_R^{\phi}} )$
 with an additional fitting parameter. 
Figure~\ref{fig:mean_square_displacement_cm-fit} (right panel)
also shows a good agreement between the values of $D_R^{\phi}$ found in these fits
and the values of the rotation diffusion coefficient $D_R(F_{\rm act},T,\phi)$ 
coming from  the late time-delay diffusive regime in the rotational MSD discussed in Sec.~\ref{sec:rotational}.  
}
 
The crossover time-delay between the last ballistic or super-diffusive, and the diffusive regimes 
increases, though rather weakly, with $\phi$, at low Pe,  see the inclined dashed line in the last panel that is also a guide-to-the-eye. 
%This crossover time-delay is the one that we could associate to the $t_C$ in~\cite{wu}.
%The strongest effect of density is on the first diffusive or sub-diffusive regime.

In~\cite{Suma14c} we showed that the qualitative dependence of  
$D_A$ on $k_BT$ for the single dumbbell case ($\phi=0$), is maintained under interactions
($\phi \neq 0$). For Pe $\ll 1$, 
$D_A$ is dominated by thermal fluctuations and increases with  $k_B T$. 
Instead, for Pe $\gg 1$, $D_A$ is dominated by the work done by the active forces.  
$D_A$ saturates at small values of $T$ for $\phi >0.2$. Instead, at high temperatures 
$D_A$ seems to retain the linear growth with temperature of the single dumbbell 
at least for the temperatures used in the simulations.

The $\phi$ dependence of $D_A$ at fixed $T$ and 
for different active forces was discussed in~\cite{Suma14b,Suma14c} where it was shown how 
the Tokuyama-Oppenheimer~\cite{Tokuyama} law of the passive system was simplified under activation. 
%to a decay that is  close to a simple exponential decay with increasing density. An improvement over this 
%description for low Pe was discussed in~\cite{Suma14c}, see below.
On the one hand, we showed that the ratio of diffusion coefficients
of the active system at finite density and single passive dumbbell $D_A(F_{\rm act},T,\phi)/D_{\rm cm}^{\rm pd}$
depends on $F_{\rm act}$ and $T$ only through Pe, as it  does for the single dumbbell. 
\comments{
In the figure we  fix $F_{\rm act}$ and we vary $T$, 
and the values Pe = $4, \ 20, \ 40, \ 66$ in each panel are obtained from three different combinations of $F_{\rm act}$ and $T$. 
 In all panels the collapse of data is very good. Note that the concavity of the 
 collapsed data changes at Pe = 20. This value is relatively far from the transition between homogeneous and segregated phases
 estimated in~\cite{Suma14,Suma14b}, and the system configurations are still homogeneous though with a not too narrow 
 distribution of local densities, $\phi_x$. 
}
The numerical data suggested~\cite{Suma14b,Suma14c}
%\begin{equation}
%\frac{D_A(F_{\rm act}, T, \phi)}{D_A(F_{\rm act}, T,0)} = \frac{f_A(\mbox{Pe}, \phi)}{f_A(\mbox{Pe}, 0)}
% \; . 
%\end{equation}
%In~\cite{Suma14b,Suma14c} we proposed
%\begin{equation}
% \frac{D_A(F_{\rm act}, T, \phi)}{D_A(F_{\rm act}, T,0)} = e^{-b(F_{\rm act}) \phi}
%\end{equation}
%Knowing now that $D_A/(k_BT)$ depends on $F_{\rm act}$ and $T$ only through  Pe, we deduce 
\begin{equation}
D_A(F_{\rm act}, T, \phi) \simeq D_{\rm cm}^{\rm pd} \ (1+\mbox{Pe}^2/8) \ e^{-b({\rm Pe}) \phi} 
\label{eq:exponential-fit}
 \end{equation}
 with $b$ a non-monotonic function of Pe~\cite{Suma14b}
% Note that in~\cite{Suma14b} the maximum in $b$ appeared at $F_{\rm act} \simeq 0.1$ that, for the temperature 
% used, $T=0.05$, corresponds to Pe $\simeq 4$.  Thus, $f_A(\mbox{Pe}, \phi)$ should
% be monotonically increasing with Pe, at all fixed $\phi$, for Pe $\geq 4$ 
% as occurs when comparing the data on the different panels in Fig.~\ref{fig:diffusion_cm}.
% In Fig.~\ref{fig:diffusion_cm} we included, with dotted black lines, the exponential fits in Eq.~(\ref{eq:exponential-fit})
% where the only free parameter is $b(\mbox{Pe})$. The values of $b(\mbox{Pe})$ are $1.1,\ 1.6,\  2.8,\ 4.1$ for 
% Pe = $4, \ 20, \ 40, 66$, in agreement with what we reported  in~\cite{Suma14b}. 
 for Pe = $40$ and Pe  = $66$, and 
\begin{equation}
D_A(F_{\rm act}, T, \phi) \simeq D_A(F_{\rm act}, T, 0) \ [1+ a_1 (\mbox{Pe}) \ \phi + a_2(\mbox{Pe}) \ \phi^2]
\label{eq:quadratic-fit}
\end{equation}
for Pe = 4 and Pe = 20, with $a_1$ negative in all 
cases while $a_2$ changing sign  from negative at Pe $<  20$ to positive at Pe $> 20$ (leading to a growing behaviour 
at large $\phi$ that is not physical). At Pe = 20 the 
density dependence is almost linear as $a_2$ is very close to zero.

\begin{figure}[!h]
\begin{center}
  \begin{tabular}{cc}
      \includegraphics[scale=0.67]{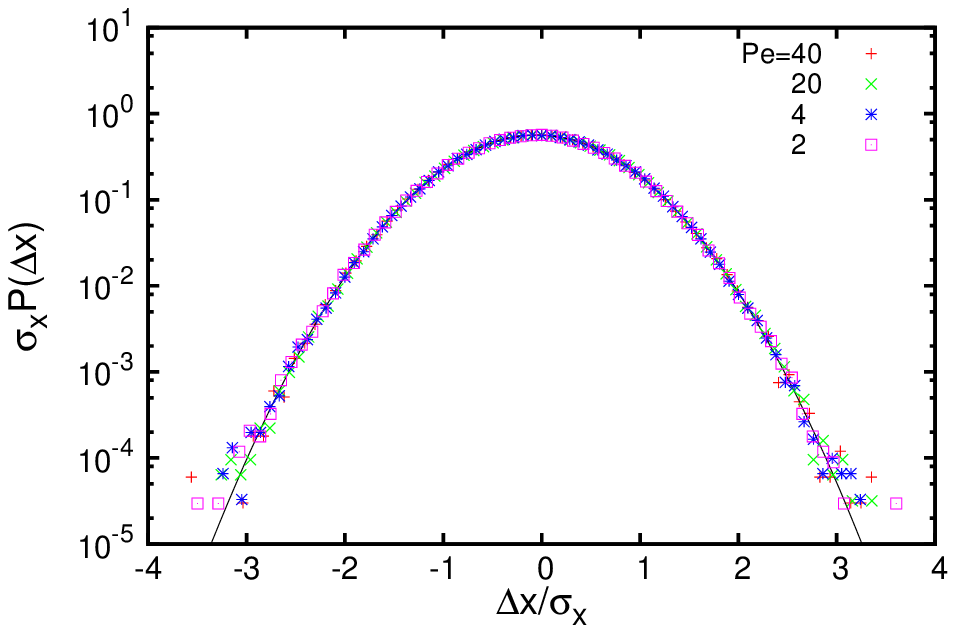}
      \includegraphics[scale=0.67]{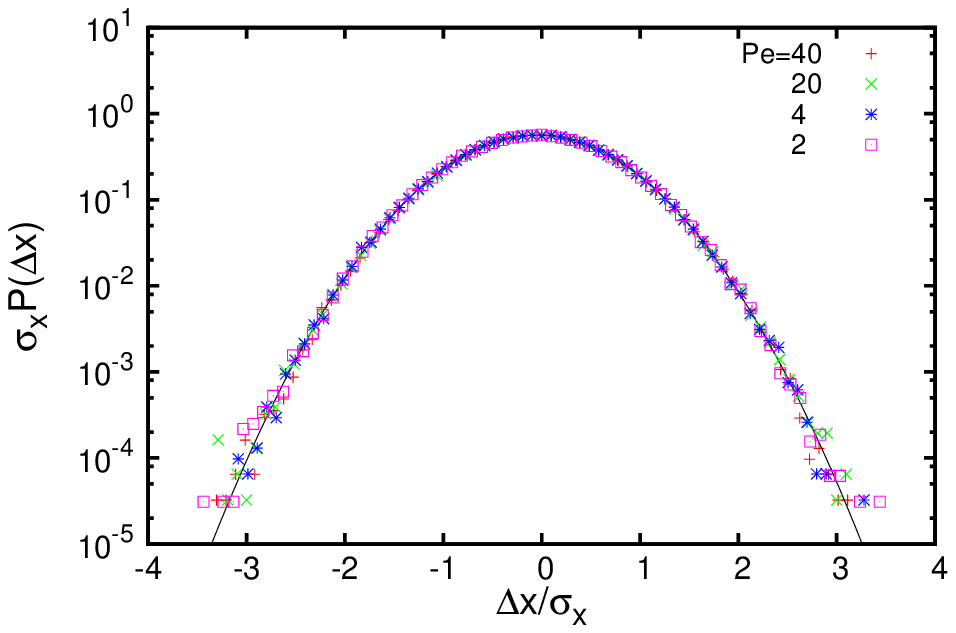}\\    
     \includegraphics[scale=0.67]{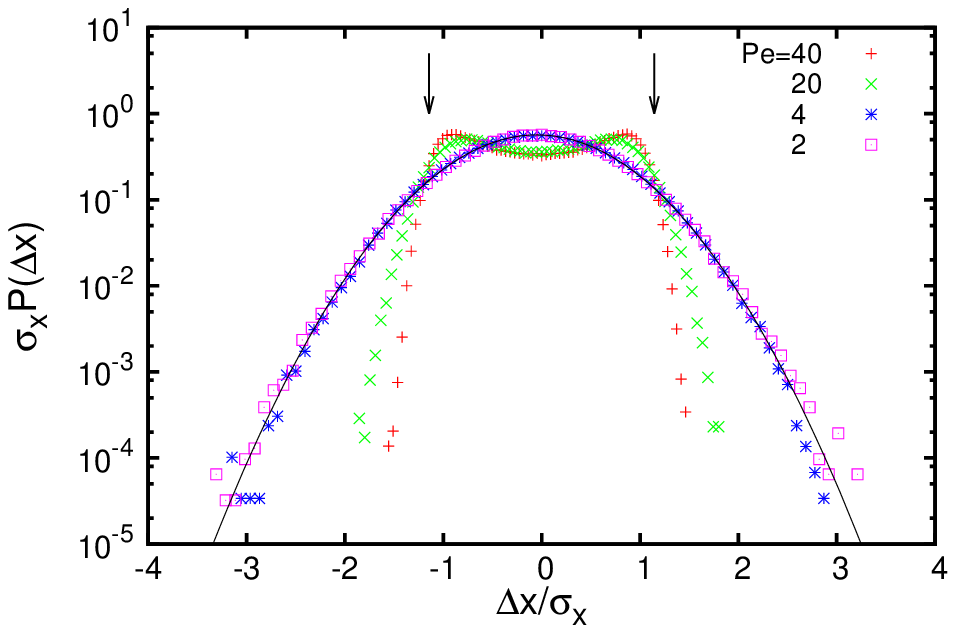}    
        \includegraphics[scale=0.67]{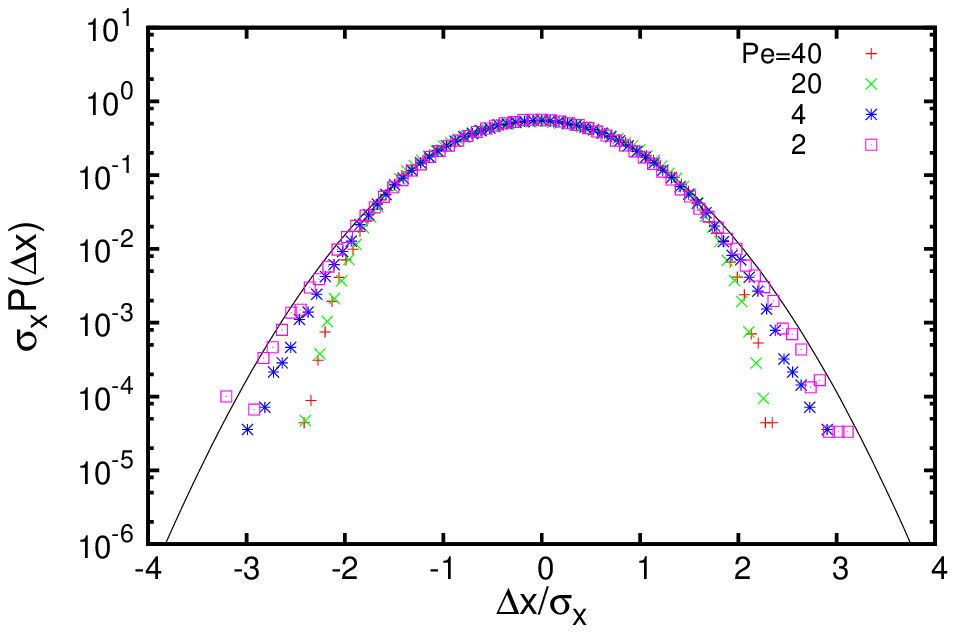}\\
    \end{tabular}
\caption{
Normalized distribution of the centre of mass horizontal displacements, $\Delta x$, in a system with 
density $\phi =0.1$. Four P\'eclet numbers are used 
in each panel,  Pe = $40, \ 20, \ 4, \ 2$ or, equivalently, $T= 0.005,\ 0.01, \ 0.05,\ 0.1$, 
as indicated in the keys. The time-delay in each of the four panels is 
$t = t_I,  \ 10 \ t_I, \ t_a/2, \ 10 \ t_a$ (from left to right and from top to bottom).
$t_I=0.1$ and the values of $t_a$ depend on Pe as given in Table~\ref{table-uno}.
The probability distribution functions are scaled with 
$\sigma_x= \langle\Delta x^2\rangle^{1/2}$. The solid curves are Gaussian pdfs in normal 
form (with zero mean and unit variance) and represent the data in the first two panels ($t = t_I, \ 10 \ t_I$) rather 
accurately.  Instead, the data for high Pe in the last two panels ($t = t_a/2, \ 10 \ t_a$) are not Gaussian.
See the main text for a discussion.
}
\label{fig:hist} 
\end{center}
\end{figure}

\subsection{The fluctuations of the centre of mass displacements}

We study the distribution of centre of mass displacements
\begin{equation}
P(\Delta x) = \frac{1}{N} \sum_{i=1}^N \langle \delta\left(\Delta x - (x_{\rm cm_i}(t+t_0) - x_{\rm cm_i}(t_0)) \right) \rangle
\end{equation}
or the self-part of the van Hove correlation function. 

In the zero density limit and under no active force the distribution of $\Delta x$ is Gaussian. We will
now determine how this limiting form is modified by the active force and the interactions in the various time-delay 
regimes.

In Fig.~\ref{fig:hist} we display the statistics of the horizontal center of mass displacements in a system 
with  a relatively low density, $\phi=0.1$. Data for Pe = $40, \ 20, \ 4, \ 2$ are gathered in each of the panels
that correspond to different time-delays, $t = t_I, \ 10 \ t_I, \ t_a/2$ and  $10 \ t_a$. The data are 
scaled by the horizontal contribution to the mean-square displacement, $\sigma_x = \langle \Delta x^2\rangle^{1/2}$.
The solid lines are Gaussian pdfs in their normal form and describe the data for short time-delays and all Pe in the 
first two panels very accurately. Instead, the data in the last two panels are close to the Gaussian form only 
for the low Pe's while for high Pe's the  shape of the pdfs is different.
In the third panel in Fig.~\ref{fig:hist} the time-delay, 
$t = t_a/2$, is such that the mean-square displacements are in the second ballistic regime, that in the single dumbbell 
case corresponds to $\sigma^2 = \langle \Delta {\mathbf r}^2_{\rm cm}\rangle 
\simeq (F_{\rm act}/\gamma)^2 \ t^2$. The distribution of $\Delta r$ is peaked around a constant value of   
$F_{\rm act} t_a/(2\gamma)$ which projected on one axis gives this double shape. 
The down-pointing arrows inside the plot indicate these instants for $Pe = 40$ and are 
%pretty
 close to the 
location of the peaks in the finite density case.
% True? ***}\textcolor{blue}{abbastanza}

\begin{figure}[!h]
\begin{center}
  \begin{tabular}{cc}
      \includegraphics[scale=0.67]{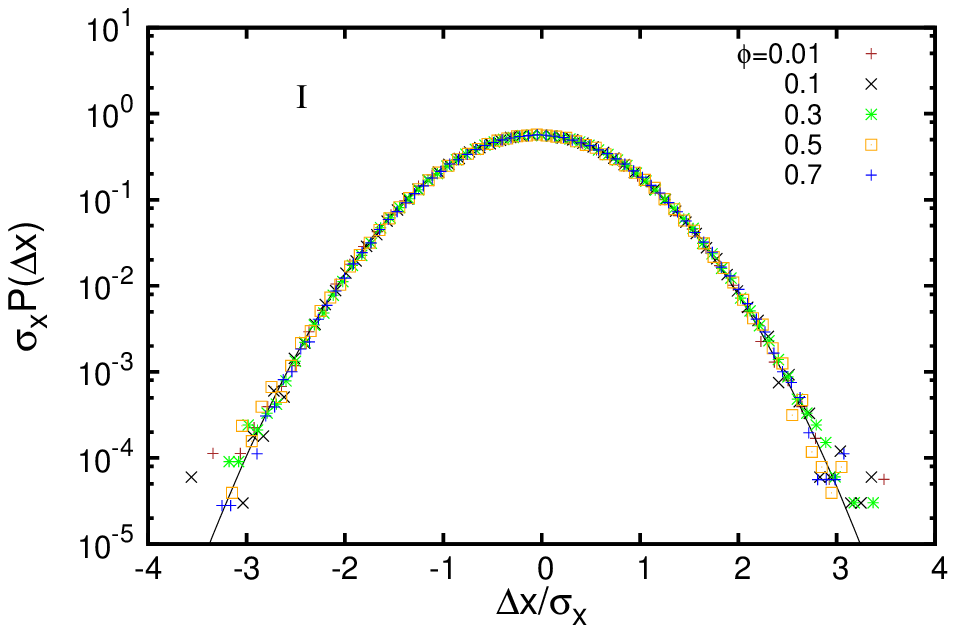}
     \includegraphics[scale=0.67]{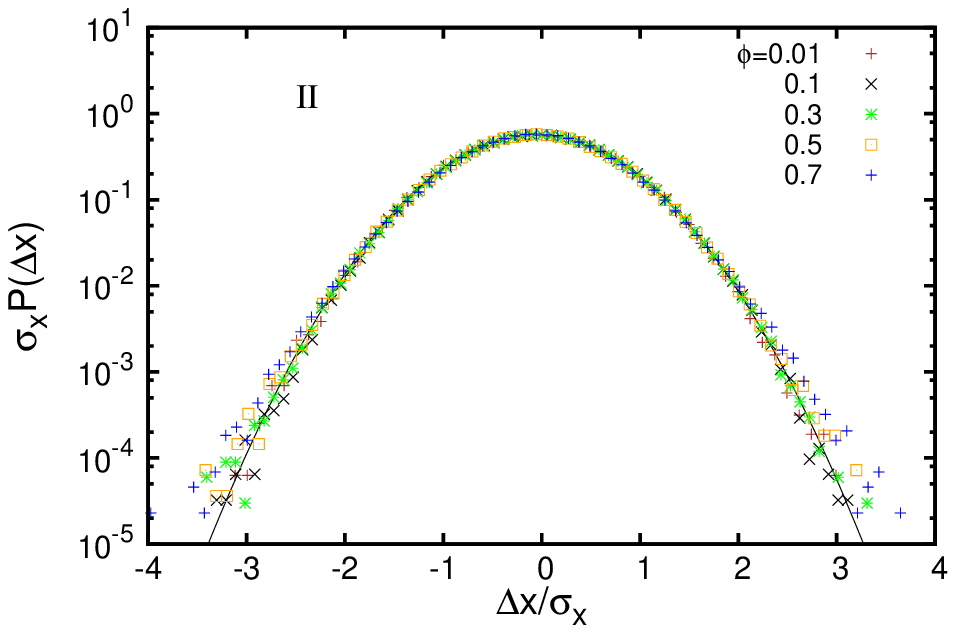}    \\
     \includegraphics[scale=0.67]{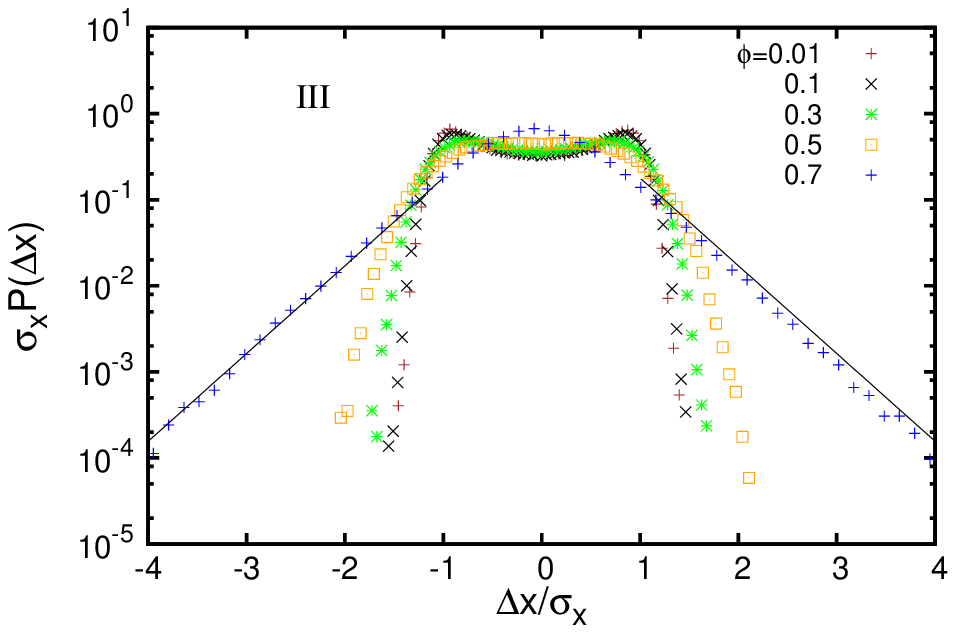}
     \includegraphics[scale=0.67]{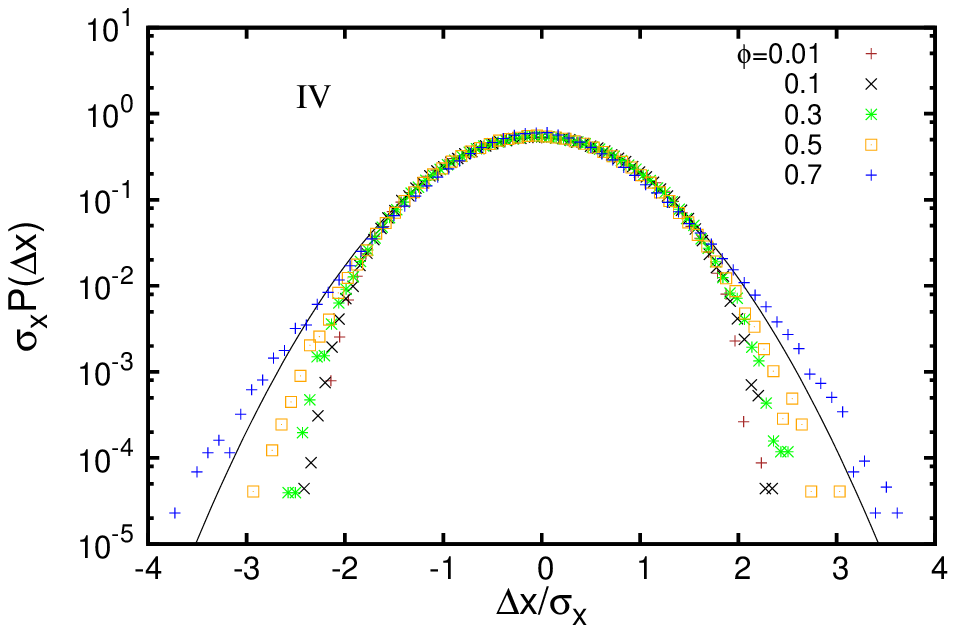}    \\
    \end{tabular}
\caption{Distribution of center of mass translation displacement in the horizontal direction, $\Delta x$,
for Pe = 40 (or $T=0.005$),  and various densities  $\phi=0.01, \ 0.1, \ 0.3, \ 0.5, \ 0.7$. The four panels
 correspond to time-delays in the four regimes labelled
 I, II, III and IV ($t_I, \ 10 \ t_I, \ t_a/2$ and $10 \ t_a$, respectively).  For I and II the Gaussian in normal form describes the data very well.
 In the last two panels the statistics are not Gaussian and there is a non-trivial density and time-delay dependence
 that we discuss in the main text.  The straight lines in the third panel ($t=t_a/2$) are exponential fits to the tails of the $\phi=0.7$ data.
 }
\label{fig:hist-Deltax-Pe40} 
\end{center}
\end{figure}

From the Gaussian fit of the bare data for $P(\Delta x)$ in the last diffusive regime ($t=10 \ t_a$, fourth panel in Fig.~\ref{fig:hist}) 
we extracted $D_A$ and we found good agreement with the values of $D_A$ obtained  in the analysis of the mean-square
displacement~\cite{Suma14c} (not shown).

\begin{figure}[!h]
\begin{center}
  \begin{tabular}{cc}
      \includegraphics[scale=0.67]{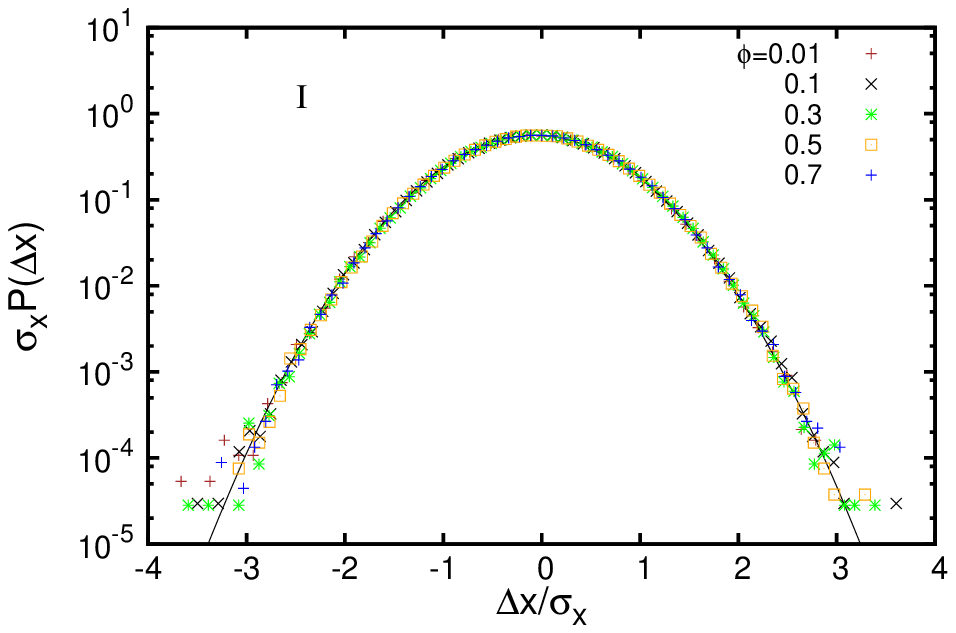}
     \includegraphics[scale=0.67]{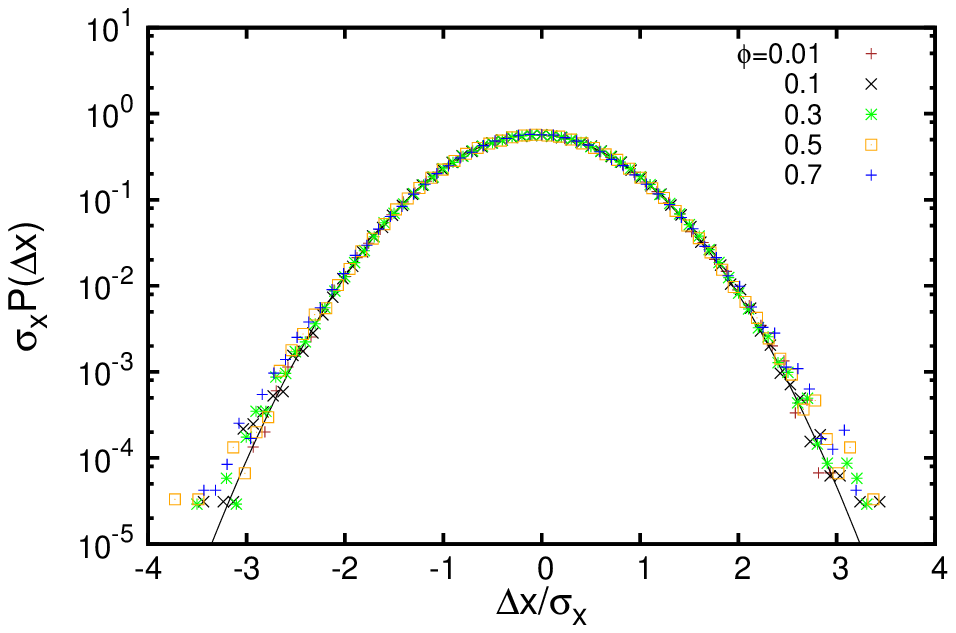}    \\
     \includegraphics[scale=0.67]{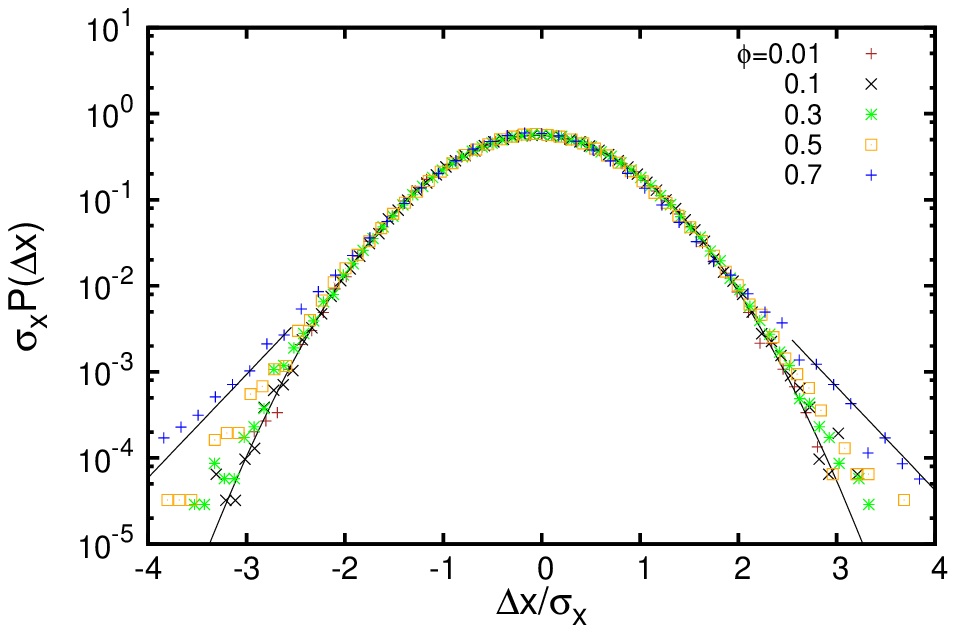}
     \includegraphics[scale=0.67]{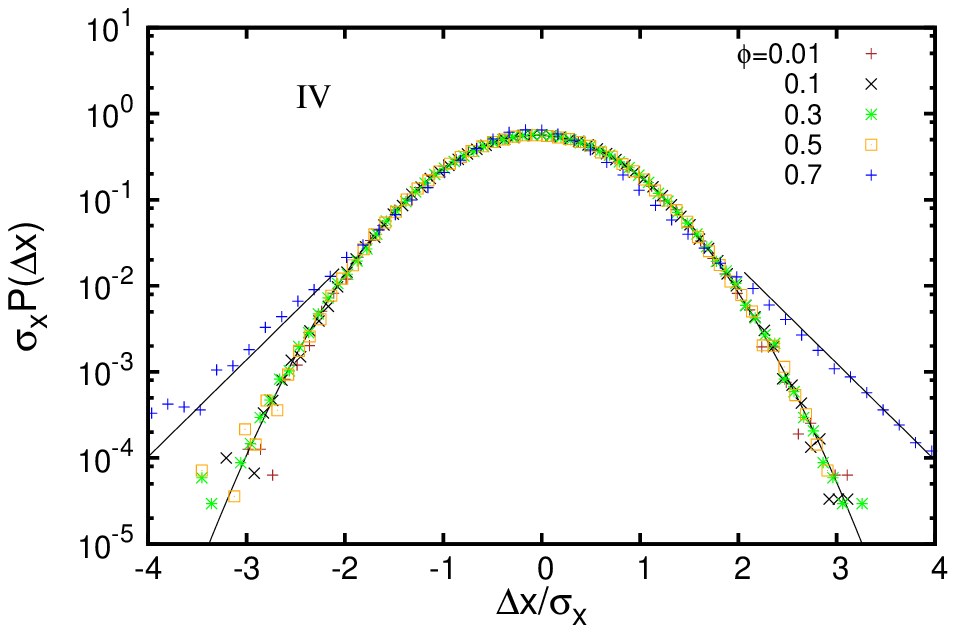}    \\
    \end{tabular}
\caption{
Distribution of the center of mass translation displacement in the horizontal direction, $\Delta x$,
 for  Pe = 2,  and various densities  $\phi=0.01, \ 0.1, \ 0.3, \ 0.5$. The four panels
 correspond to time-delays $ t=t_I, \ 10 \ t_I, \ t_a/2$ and $10 \ t_a$, respectively. The first and last cases are in the regimes labelled
 I and IV for $\phi \leq 0.5$.  As in Fig.~\ref{fig:hist-Deltax-Pe40} the tails of the distribution for $\phi=0.7$, and $t= t_a/2$ and $t=10 \ t_a$ in this case, 
 are well described by exponential fits, represented as straight lines. See the main text for a discussion.  }
\label{fig:hist-Deltax-Pe2} 
\end{center}
\end{figure}

In the following two figures, Figs.~\ref{fig:hist-Deltax-Pe40} and \ref{fig:hist-Deltax-Pe2}, we checked the density dependence of the
centre of mass displacement distribution. From the first to the fourth panels $t=t_I, 10 \ t_I, \ t_a/2, \ 10 \ t_a$, respectively. In 
each panel data for $\phi = 0.01, \ 0.1, \ 0.3, \ 0.5, \ 0.7$ are shown. 

In the cases in which activation is strong, Fig.~\ref{fig:hist-Deltax-Pe40}, at short time delays (first two panels) 
all systems have Gaussian fluctuations. At long-time delays (last two panels) the distributions depend on $\phi$. 
In the regime III 
there is a two peak structure at low density, while the central part becomes close to flat at higher densities,
due to stronger interactions, and then develops two exponential wings at larger absolute displacements. 
In the regime IV the distributions do not have the double peak structure and are
Gaussian close to $\Delta x \simeq 0$ for all densities.
Looking carefully at the MSD for Pe = 40 in the left panel in Fig.~\ref{fig:mean_square_displacement_cm}, one notices 
that the curves in this regime are less flat at low density while they are closer to being flat at $\phi=0.5$. The 
crossover to normal diffusion is, therefore, slower for lower density. The long-lasting super-diffusive behaviour 
seems to be related to the non-Gaussianity of the pdfs. The MSD for Pe = 2 in the right panel in 
Fig.~\ref{fig:mean_square_displacement_cm} are flatter earlier for $\phi \leq 0.5$ and, consistently, the 
pdfs are Gaussian for all these densities (see Fig.~\ref{fig:hist-Deltax-Pe2}). 
The case $\phi=0.7$ is different since the system is far from the last 
diffusive regime for the time-delays used.

In the cases in which activation is weak, Fig.~\ref{fig:hist-Deltax-Pe2}, once again at short time delays (first two panels) 
all systems have Gaussian fluctuations. At the intermediate time delays (third panel) the distributions depend on $\phi$ and, 
for high densities, see $\phi=0.5, 0.7$ in the plot, there is an excess weight on large deviations with respect to the Gaussian, 
that is close to exponential.
The  effect becomes more evident when the  sub-diffusive behavior is more pronounced, see
 Fig.~\ref{fig:mean_square_displacement_cm}. 
At still longer time-delays (fourth panel) the statistics becomes Gaussian again for all densities, except for the case 
at $\phi=0.7$ since in this case the time-delay is still too short to reach the final diffusive regime.

\section{Rotational motion at finite density}
\label{sec:rotational}

\begin{figure}[!h]
\begin{center}
  \begin{tabular}{cc}
      \includegraphics[scale=0.67]{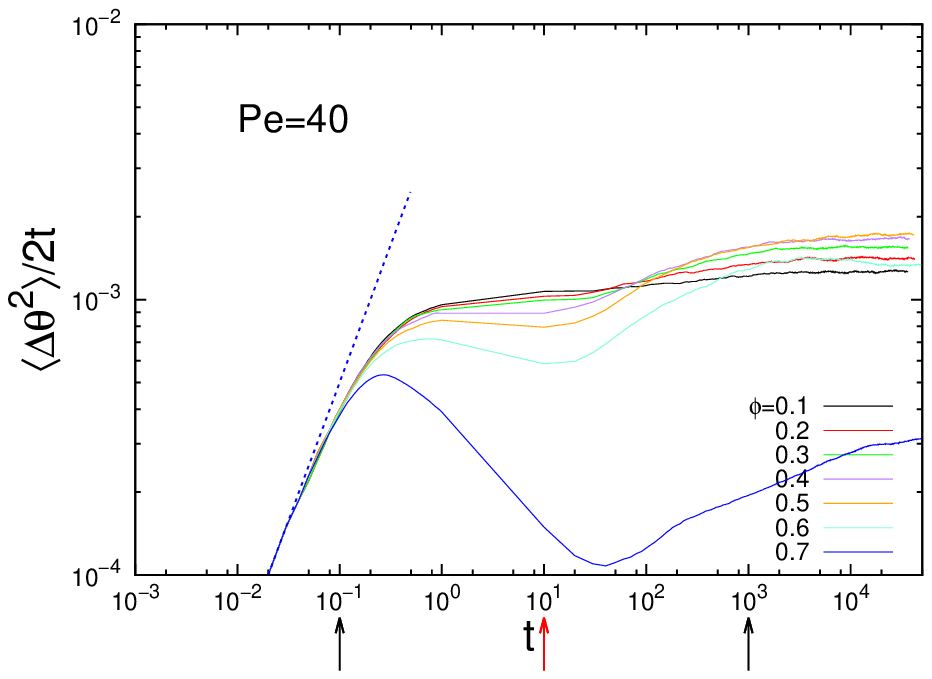}
          \includegraphics[scale=0.67]{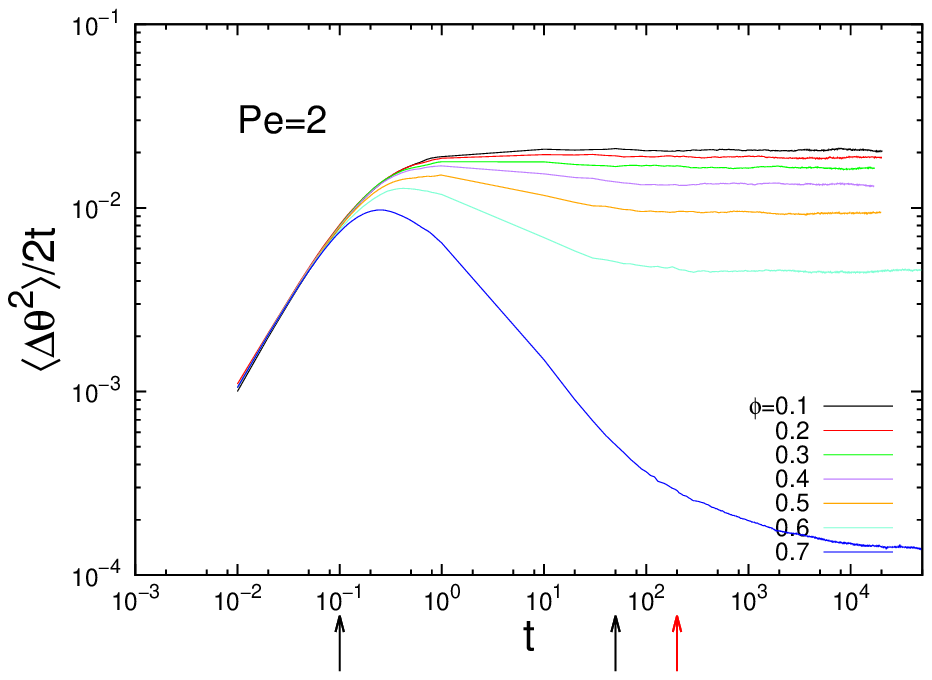}    \\
    \end{tabular}
\caption{The angular MSD for Pe = 40 and Pe = 2  and, with different lines, various 
densities given in the keys to each panel.
The dash in the first panel highlights the initial ballistic behavior. The vertical arrows
indicate the characteristic times $t_I, \ t^*,\  t_a$.  Note that the vertical scale is different in the two
panels.
}
\label{fig:mean_square_displacement_angular} 
\end{center}
\end{figure}

We turn now to the rotational dynamics.
In Fig.~\ref{fig:mean_square_displacement_angular}  we display  the angular MSD normalized by time-delay. 
The two panels show data for $F_{\rm act}=0.1$ and Pe = $40$ and Pe = 2, 
and the same  densities as in Fig.~\ref{fig:mean_square_displacement_cm}.
These plots also show interesting features:\\
\noindent
-- In all cases there is a first ballistic regime with a pre-factor that is independent of $\phi$ and increases with temperature
(The case $t\ll t_I=m_{\rm d}/\gamma$ of the single dumbbell.)
\\
-- Next,  the dynamics slow down and, depending on $T$ and $\phi$, the normalised MSD 
may attain an ever-lasting plateau associated to normal diffusion  for low $\phi$ at any temperature,
or even decrease, suggesting sub-diffusion,
at high enough $\phi$. 
\\
-- At high Pe  
and sufficiently high density the dynamics accelerate next, with a second super-diffusive regime
that crosses over to a final diffusive regime.
\\
-- In the late normal diffusive regime  all curves saturate and the height of the  plateau yields the 
different $D_R$ coefficients that we discuss below.

In the phase separated regime the dumbbell clusters rotate~\cite{Suma13,Suma14}.
It is possible that strong fluctuations not far from the critical point
have an important rotational component than enhances/advects rotational 
diffusion giving rise to an observable contribution to displacement.

The diffusion constant decreases with temperature below a crossover  beyond which it
increases approximately linearly with temperature~\cite{Suma14c}. 
The $F_{\rm act}$-independence of $D_R$ is lost as soon 
as the interaction between dumbbells is switched on. The density dependence is also 
quite complex and was discussed in~\cite{Suma14c}.
Finally, we showed that $D_R/(k_BT)$ depends on $F_{\rm act}$ and $k_BT$ only through Pe,
%we analyse whether $D_R/(k_BT)$ depends  on $F_{\rm act}$ only via Pe. To this end, in Fig.~\ref{fig:diffusion_angular_Pe}
\begin{equation}
D_R(F_{\rm act}, T, \phi) = k_BT \ f_R(\mbox{Pe}, \phi)
\end{equation}
with $f_R(\mbox{Pe},0) = f_R(0,0) = 2/(\gamma \sigma_{\rm d}^2)$. 
At low densities, 
while the master curve decreases with  $\phi$ in the whole range considered for Pe $< 20$, 
it becomes flat at Pe $=20$ and it increases with $\phi$ for Pe $> 20$,  for $\phi \le 0.5$. 
This would suggest:
\begin{equation}
f_R (\mbox{Pe},\phi) \simeq \frac{2}{\gamma \sigma_{\rm d}^2 } + a(\mbox{Pe},\phi)  \ , 
\end{equation}
with $a($Pe, $\phi)$ almost linear in $\phi$ and the slope changing sign at Pe $\simeq 20$ for small $\phi$. At 
all Pe there is a cross-over at high enough densities after which the rotational diffusion constant 
decreases with increasing density. One can associate this feature to the fact that for
sufficiently dense systems rotations are inhibited and $D_R$ decreases. 

\subsection{The angular displacement distribution}

\begin{figure}[!h]
\begin{center}
  \begin{tabular}{cc}
      \includegraphics[scale=0.67]{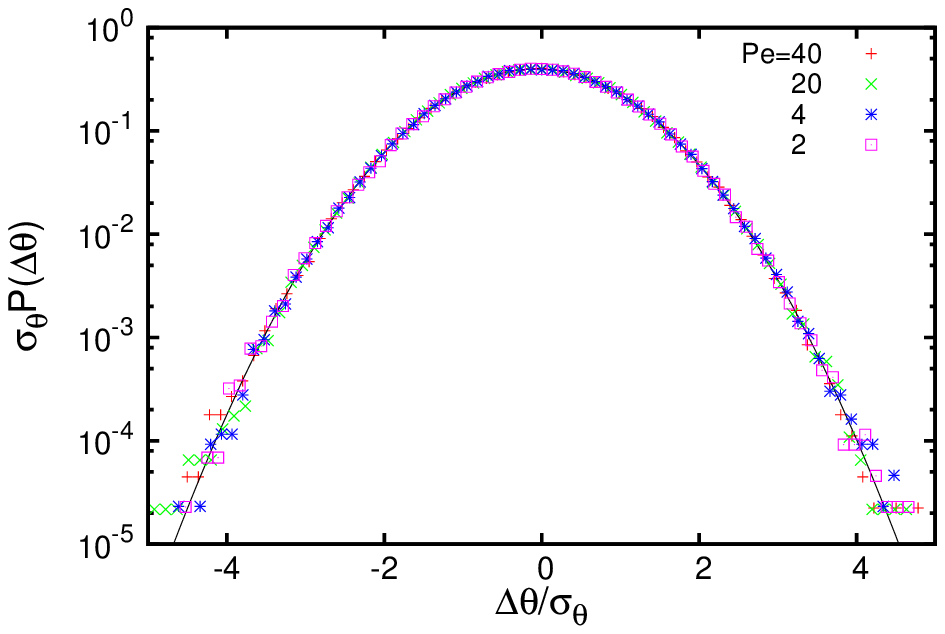}
     \includegraphics[scale=0.67]{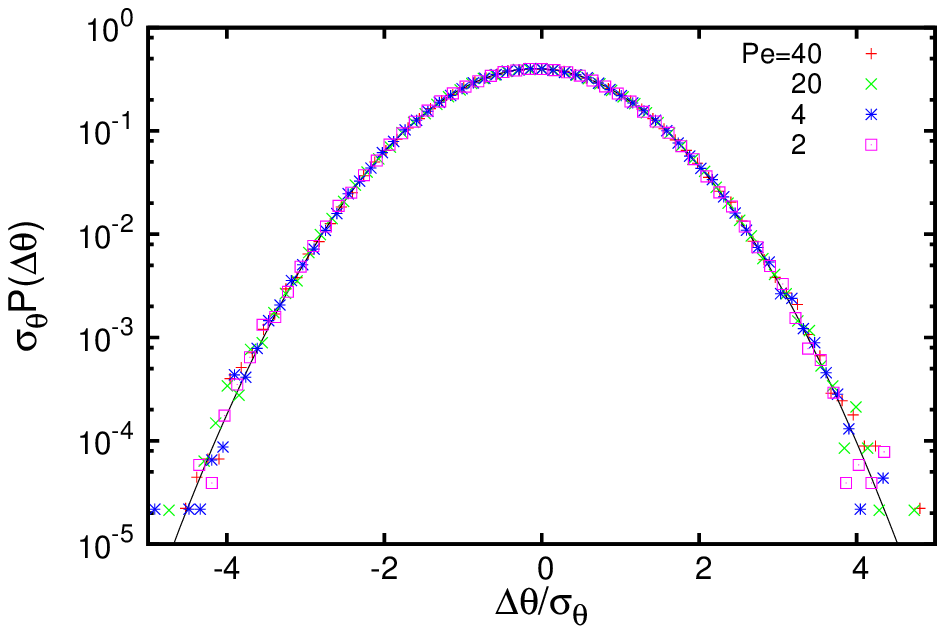}    \\
     \includegraphics[scale=0.67]{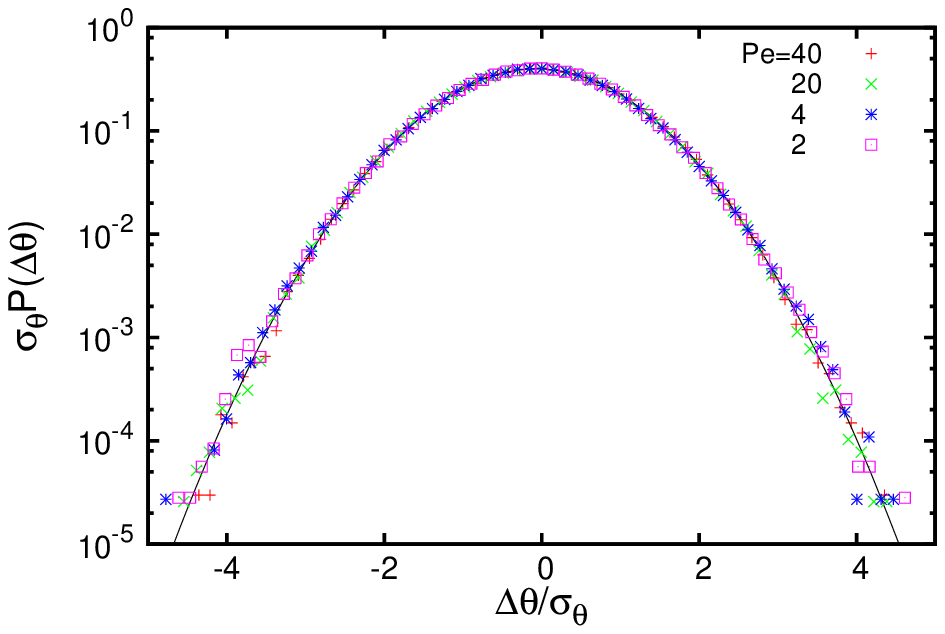}    
        \includegraphics[scale=0.67]{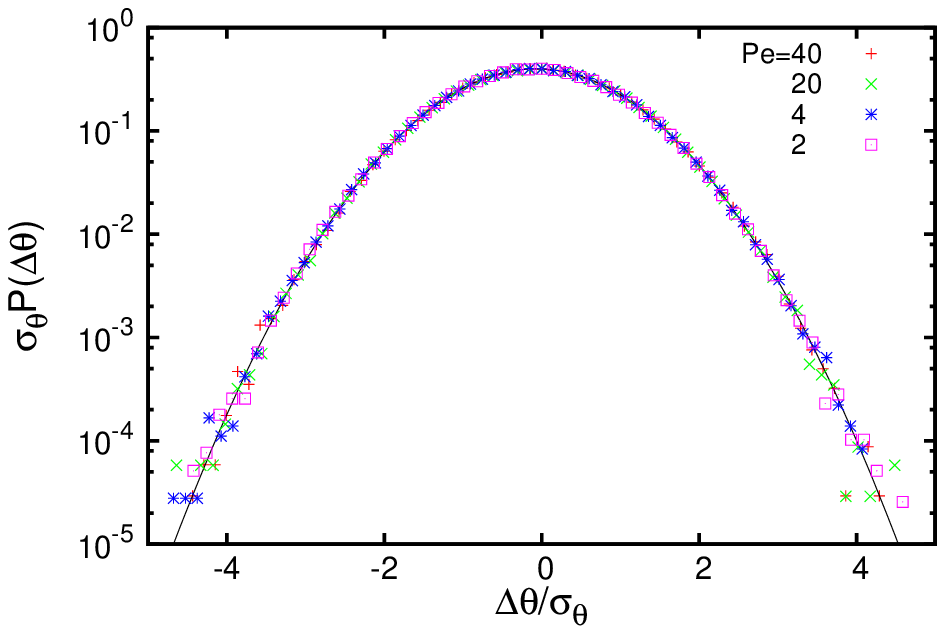} \\
    \end{tabular}
\caption{Normalized distribution of angular displacements, $\Delta \theta$, in a low-density system,
$\phi=0.1$. Four P\'eclet numbers are used 
in each panel,  Pe = $40, \ 20, \ 4, \ 2$, 
as indicated in the keys. The time-delays in the four panels are 
$t = t_I,  \ 10 \ t_I, \ t_a/2, \ 10 \ t_a$. 
The values of $t_a$ depend on Pe and are given in Table~\ref{table-uno}.
The probability distribution functions are scaled with 
$\sigma_\theta= \langle\Delta\theta^2\rangle^{1/2}$ and the solid lines are a Gaussian in normal 
form, with zero mean and unit variance. 
}
\label{fig:hist-theta-Pe-phi-0.1} 
\end{center}
\end{figure}

Figure~\ref{fig:hist-theta-Pe-phi-0.1} demonstrates that for low enough density, $\phi=0.1$ in this figure,  
the angular displacement 
distribution function at all time delays and for all P\'eclet numbers 
can be put into a normal Gaussian form after normalisation by 
$\sigma_\theta = \langle \Delta\theta^2\rangle^{1/2}$. 

\begin{figure}[!h]
\begin{center}
  \begin{tabular}{cc}
      \includegraphics[scale=0.67]{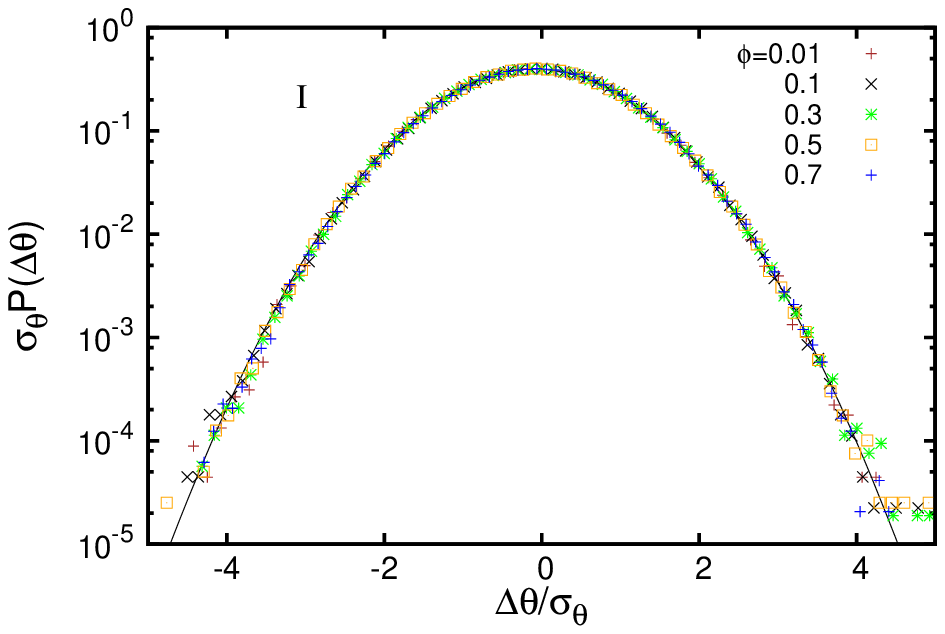}
     \includegraphics[scale=0.67]{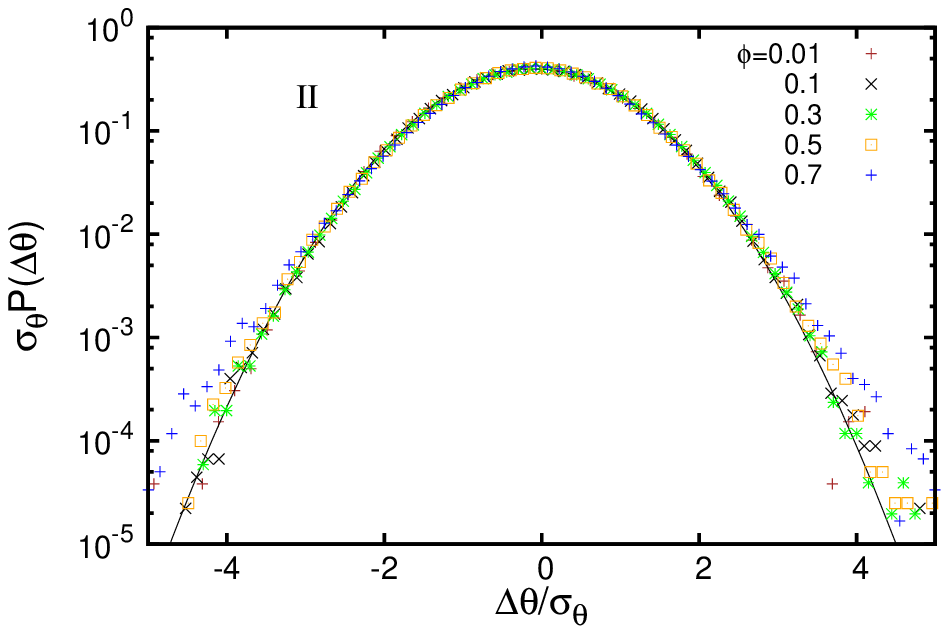}    \\
     \includegraphics[scale=0.67]{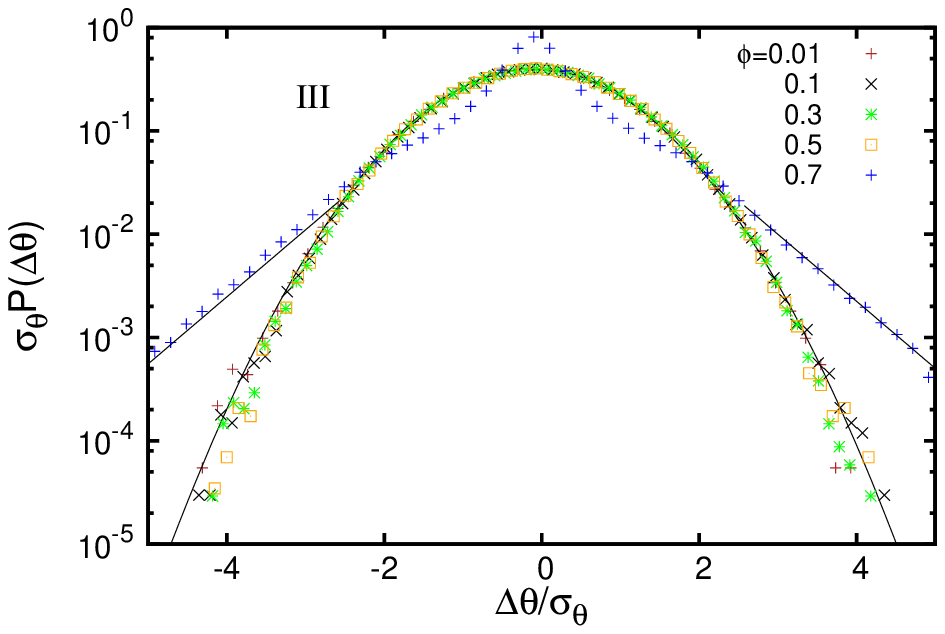}
     \includegraphics[scale=0.67]{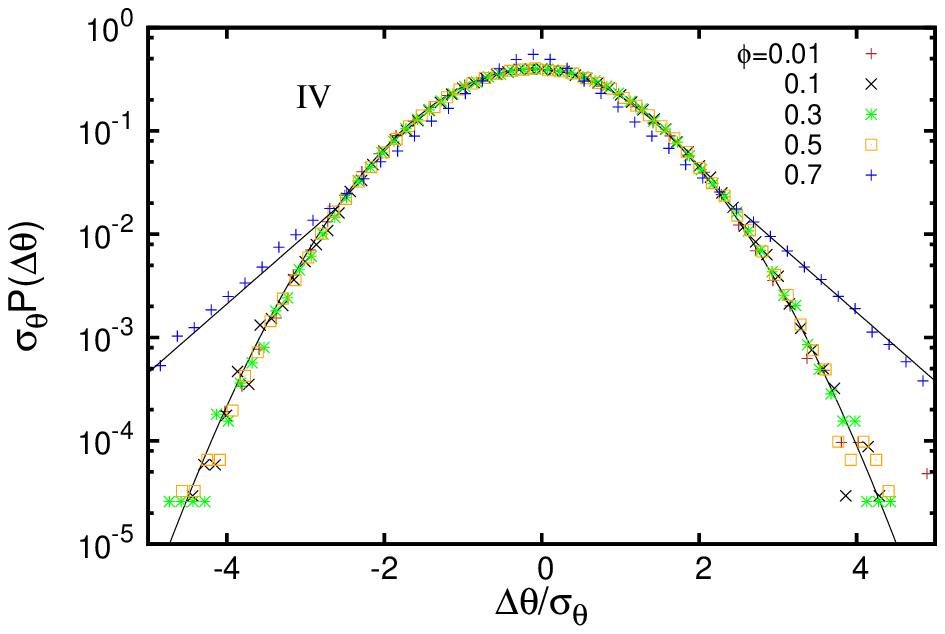}    \\
    \end{tabular}
\caption{Normalized distribution of angular displacements $\Delta \theta$ for Pe = $40$. Data for 
various densities $\phi=0.01,\ 0.1, \ 0.3, \ 0.5, \ 0.7$ are collected in each panel. 
In each of the four panels the time-delay ($t = t_I, \ 10 \ t_I, \ t_a/2,\ 10 \ t_a$) is chosen to lie 
in one of the four relevant time-scales called I, II, III, IV. $t^* = 10 \ll t_a =1000$,
see Table~\ref{table-uno}, for this high value of Pe. The solid curve is a Gaussian in normal form.
In the second, third and fourth panels deviations from the Gaussian are visible for the high density data, $\phi=0.7$.
In the last two panels the straight lines are exponential fits to the tails.}
\label{fig:hist-theta-Pe40-phi} 
\end{center}
\end{figure}

\begin{figure}[!h]
\begin{center}
  \begin{tabular}{cc}
      \includegraphics[scale=0.67]{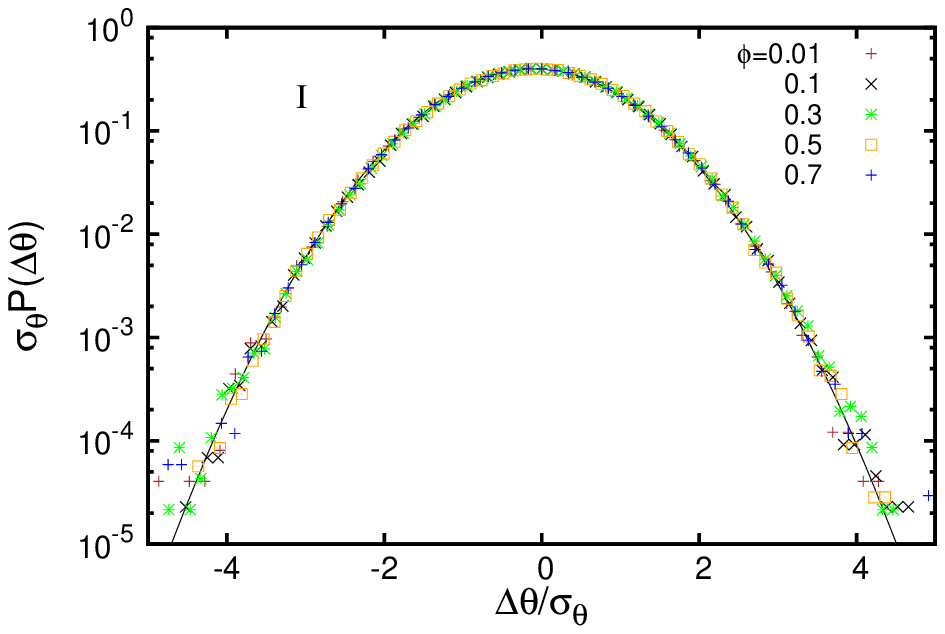}
     \includegraphics[scale=0.67]{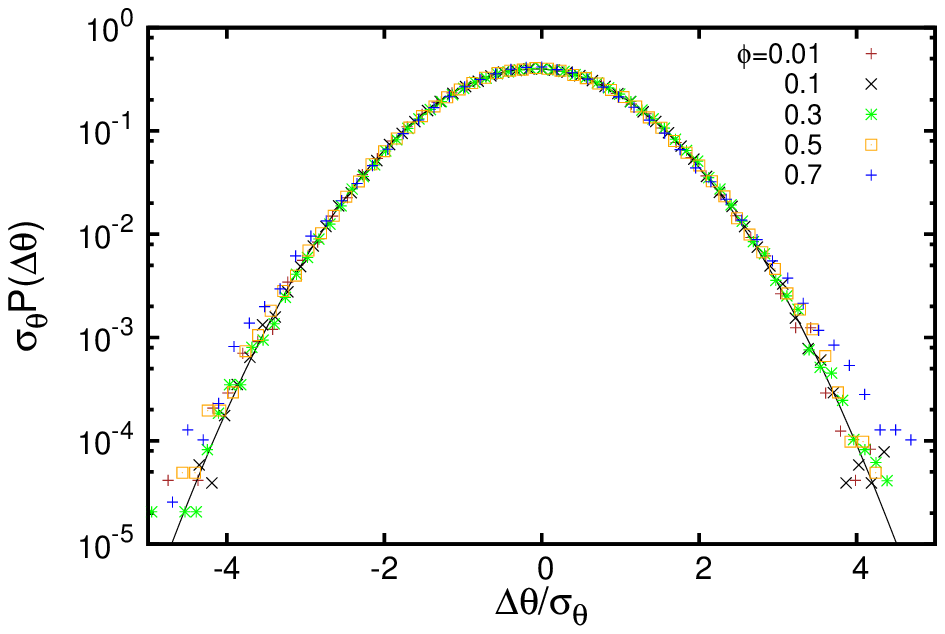}    \\
     \includegraphics[scale=0.67]{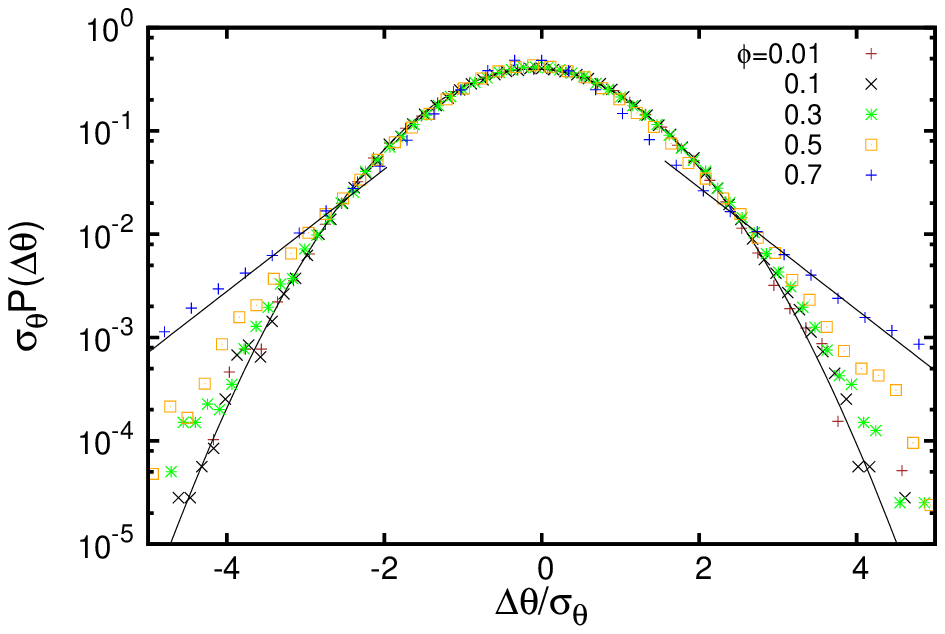}
     \includegraphics[scale=0.67]{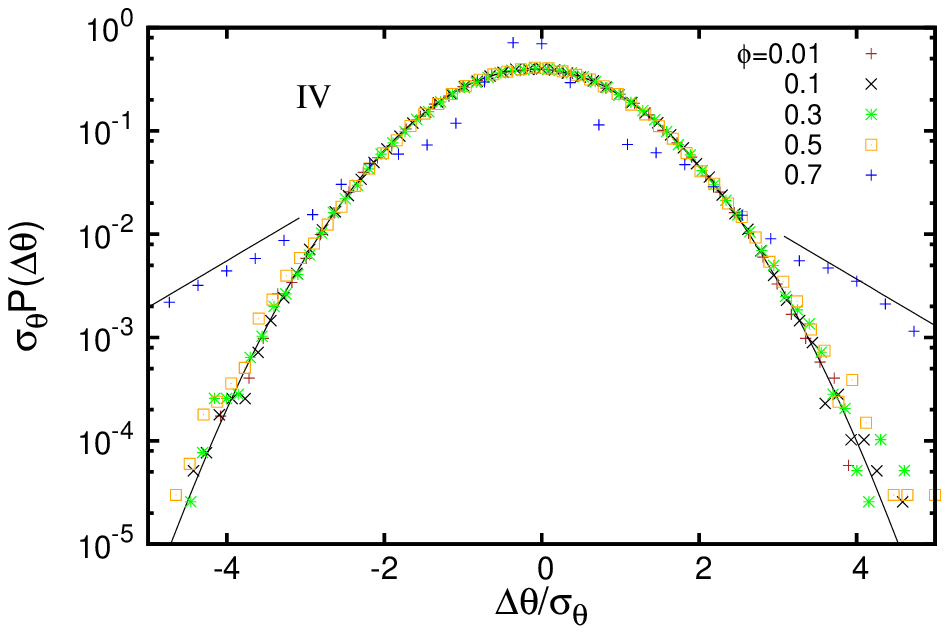}    \\
    \end{tabular}
\caption{
Normalized distribution of angular displacements $\Delta \theta$ for Pe = $2$. Data for various densities 
$\phi=0.01,\ 0.1, \ 0.3, \ 0.5, \ 0.7$ are collected in each panel. The  time-delay in the first ($t=t_I$) and last ($t=10 \ t_a$) panels lie in 
the inertial ballistic and last diffusive regimes I and IV, respectively, for $\phi\leq 0.5$. The time-delays in the second and third panels 
($t=10\ t_I$ and $t=t_a/2$, respectively) are in the intermediate crossover between these two. 
$t^* = 4000 \gg t_a = 50$, see the Table~\ref{table-uno}, for this low Pe number. The solid curves are 
Gaussians in normal form. The deviations from the Gaussian in the intermediate regime are 
pronounced for lower densities than for Pe = 40. The solid straight lines are exponential fits to the tails in the 
$\phi=0.7$ data.}
\label{fig:hist-theta-Pe2-phi} 
\end{center}
\end{figure}

In Figs.~\ref{fig:hist-theta-Pe40-phi} and \ref{fig:hist-theta-Pe2-phi} we analyse the density dependence of the 
normalised probability distribution of angular displacements. In Fig.~\ref{fig:hist-theta-Pe40-phi} a 
high P\'eclet number is used, Pe = 40, while in Fig.~\ref{fig:hist-theta-Pe2-phi} the 
P\'eclet number is low, Pe = 2. In the first case, the four time-regimes I, II, III, IV, in the behaviour of the
center of mass translational mean-square displacement, are well separated. In the second case, they are
not. We see that for $\phi=0.01, \ 0.1, \ 0.3$ the data in both figures collapse onto the normal 
Gaussian.  The higher density data, $\phi=0.5$, deviate from this master curve when the 
time delay is chosen to be $t=10 \ t_I$ for Pe = 40, and $t = t_a/2 =25$ for Pe = 2. The data points 
for $\phi=0.7$ in the last two panels in the two figures have a peak at $\Delta\theta=0$ corresponding to 
dumbbells that do not rotate  between the two times and two exponential wings typical of heterogeneous 
systems. Note that these pdfs resemble the ones in~\cite{Levis-Berthier} though for translational displacements in this case.

\section{Conclusions}

We  considered the diffusion properties of a system of  active dumbbells with repulsive interaction.  
We focused on the P\'eclet number and density regime in which the global 
system is homogeneous and we studied the fluctuations of the 
centre of mass translation and angular displacements.
%We reach higher densities than the ones used in~\cite{wu,Leptos09}.
%We restrain the Pe to be small enough (and possibly much smaller than in the experiments) 
%to keep the system in its homogenous phase. 

%\textcolor{green}{Ho accorciato, ma non troppo. Possiamo anche togliere il comportamento dei coeff di diffusione.}
We first summarised the translational and rotational MSD of the system.
In the single particle  limit, the translational MSD has a very rich time-delay dependence, with 
four distinct time regimes (ballistic, diffusive, ballistic and diffusive) separated by three characteristic times
(the shortest inertial, $t_I$, the  diffusive, $t_a$, and an additional one, $t^* \propto t_a/\mbox{Pe}^2$, that lies in 
between the other two for large Pe). This rich structure survives under finite densities with modified parameters.
The diffusion constant in the last diffusive regime has a non-monotonic dependence on 
temperature, as for the single dumbbell case, and it decreases with increasing self-propelled particle
density at all temperatures. Moreover,  it depends on temperature and active force only 
through the P\'eclet number at all densities explored. In general,  the Pe dependence is non-monotonic.

The behavior of the rotational MSD, that is rather simple for a single dumbbell with just one 
crossover between ballistic and diffusive behaviour,  reflects  the behavior ot the translational MSD at finite densities,
 where intermediate regimes also appear.
 The late epochs rotational diffusion constant increases with temperature (though not linearly) at all 
densities and active forces simulated and depends on temperature 
and activity only through Pe. At low densities,
 its dependence on density changes from decreasing at low Pe to increasing at high Pe.

%The time-delay dependence of the single dumbbell rotational MSD  
%is rather simple, with a single crossover between ballistic and diffusive behaviour. Intermediate regimes appear at 
%finite densities.
%The late epochs diffusion constant increases with temperature (though not linearly) at all 
%densities and active forces simulated. The independence on active force is lost at finite densities. We proved
%that the angular diffusion constant depends on temperature and activity only through the P\'eclet number 
%and that, at low densities, its dependence on density changes from decreasing at low Pe to increasing at high Pe. We can 
%ascribe the change in behaviour to the large scale density fluctuations that appear close to the transition 
%from homogeneous to phase separated phases at a critical Pe.
%% The large and rather compact clusters
%in this region of parameters rotate coherently~\cite{Suma13,Suma14} and may be the cause for the 
%increase of $D_R$ with $\phi$. On the other hand, at large enough densities
%rotations are strongly inhibited and the value of $D_R$ decreases for all Pe. 

We have then evaluated the distribution  functions for the centre of mass and angular displacements  
for time-delays corresponding to the various dynamical regimes.  For time-delays shorter or of the order
of the inertial time $t_I$ the distributions were always found  to be very close to  a Gaussian, for all the $\phi$ and Pe 
considered. 
At large Pe,  always in the homogeneous phase  but not far from the critical value (Pe/${\rm Pe}_c \approx 0.62$)~\cite{Suma14c},
 for time-delays corresponding to the super-diffusive or second ballistic regime of the single particle,
while the angular distributions  remain Gaussian except at the highest density considered $\phi=0.7$, 
the translational displacement distributions show a two-peak character  at low density becoming more rounded when the density increases. The position of the peaks corresponds
to the analytic estimate given in Sec.~\ref{subsec:distribution-dumbbell}. 
The distributions get closer  to a  Gaussian  
 in the last regime. The effects on the final diffusive regime due to the cross-over with   the preceding  super-diffusive regime 
are more pronounced  than at small Pe even at  times of the order of  $10 \ t_a$. Accordingly, non-Gaussian effects can be seen in the translational distributions at these times.  
The case at $\phi=0.7$ is different since, for the time-delay considered, the system is still far from the final diffusive regime.
	At small Pe, when the effects of  interactions are less relevant (Pe/${\rm Pe}_c \approx 0.03$), the second regime becomes 
	sub-diffusive when the density increases.
	In correspondence with this, deviations from the Gaussian are found in both translational and angular displacement distributions. Finally, both
	distributions have a Gaussian character in correspondence with the last diffusive regime, except at $\phi=0.7$. In this very high density
	limit we find exponential tails in the translational and rotational distributions, and a central peak at vanishing angular displacement.

After this work we plan to analyse the motion of tracers in contact with this active sample and, especially,
to analyse the existence of a parameter to be interpreted as an effective temperature~\cite{cugl:review} from the 
mobility and diffusive properties of the sample and the tracers, as done 
in~\cite{cugl-mossa1,cugl-mossa2,cugl-mossa3} for another active matter model  and in~\cite{Palacci10}
for a sample of active Janus particles. 

\vspace{0.5cm}

\noindent \underline{Acknowledgments}:
LFC is a member of Institut Universitaire de France.  G.G. acknowledges the support of  
MIUR (project PRIN 2012NNRKAF).

\vspace{1cm}

\bibliographystyle{elsarticle-num} 
\bibliography{dumbbells-biblio}

\end{document}